\documentclass[journal]{IEEEtran}
\ifCLASSINFOpdf
\else
\fi
\usepackage{url}
\usepackage{tabularx} 

\usepackage[dvipdfm]{graphicx}
\usepackage{graphicx}
\usepackage{amssymb}
\usepackage{amsmath}
\usepackage[table]{xcolor}
\usepackage{comment}

\usepackage{amsfonts}
\usepackage{scalefnt}
\usepackage{color}
\usepackage{cite}
\usepackage{multicol}
\usepackage{theorem}
\usepackage{amsmath}
\usepackage{comment}
\usepackage{float}
\usepackage[caption=false]{subfig}

\usepackage[linesnumbered,ruled]{algorithm2e}

\usepackage[dvipsnames]{xcolor}
\usepackage{soul}
\usepackage{booktabs}
\usepackage{longtable}

\usepackage[acronym,nonumberlist,style=superheader,toc,nopostdot]{glossaries}

\newacronym{5g}{5G}{fifth generation}
\newacronym{4g}{4G}{fourth generation}
\newacronym{6g}{6G}{sixth generation}
\newacronym{abep}{ABEP}{average bit error probability}
\newacronym{aber}{ABER}{average bit error rate}
\newacronym{af}{AF}{amplify-and-forward}
\newacronym{awgn}{AWGN}{additive white Gaussian noise}
\newacronym{ask}{ASK}{amplitude shift keying}
\newacronym{ap}{AP}{access point}
\newacronym{aead}{AEAD}{authenticated encryption with associated data}
\newacronym{an}{AN}{artificial noise}
\newacronym{ad}{AD}{associated data}

\newacronym{bask}{BASK}{binary amplitude shift keying}
\newacronym{ber}{BER}{bit error rate}
\newacronym{bep}{BEP}{bit error probability}
\newacronym{bler}{BLER}{block error rate}
\newacronym{bpsk}{BPSK}{binary phase shift keying}
\newacronym{bs}{BS}{base station}

\newacronym{cc}{CC}{cooperative communication}
\newacronym{cv}{CV}{communication vehicle}
\newacronym{cdf}{CDF}{cumulative distribution function}
\newacronym{clt}{CLT}{central limit theorem}
\newacronym{cm}{CM}{constant modulus}
\newacronym{cfr}{CFR}{channel frequency response}
\newacronym{cp}{CP}{cyclic prefix}
\newacronym{cpi}{CPI}{codeword position index}
\newacronym{cr}{CR}{cognitive radio}
\newacronym{csi}{CSI}{channel state information}
\newacronym{crs}{CRS}{cooperative relaying system}
\newacronym{cdma}{CDMA}{code division multiple access}
\newacronym{cr-noma}{CR-NOMA}{cognitive-radio inspired NOMA}
\newacronym{ce}{CE}{channel estimation}
\newacronym{cod}{CoD}{communications device}
\newacronym{crlb}{CRLB}{Cram\'{e} r--Rao lower bounds}
\newacronym{dc}{DC}{direct current}
\newacronym{df}{DF}{decode-and-forward}
\newacronym{dl}{DL}{downlink}
\newacronym{dft}{DFT}{discrete Fourier transform}
\newacronym{dnn}{DNN}{deep neural network}
\newacronym{dwt}{DWT}{discrete wavelet transform}  
\newacronym{drl}{DRL}{deep reinforcement
learning}
\newacronym{dhke}{DH-KE}{Diffie-Hellman key exchange}

\newacronym{ec}{EC}{ergodic capacity}
\newacronym{ee}{EE}{energy efficiency}
\newacronym{eh}{EH}{energy harvesting}
\newacronym{ed}{ED}{Euclidean distance}
\newacronym{esprit}{ESPRIT}{estimation of signal parameters via rotational invariance techniques}

\newacronym{fd}{FD}{full-duplex}
\newacronym{fft}{FFT}{fast Fourier transform}
\newacronym{fec}{FEC}{forward error correcting}
\newacronym{fso}{FSO}{free space optical}
\newacronym{fe}{FE}{fuzzy extractor}
\newacronym{fdma}{FDMA}{frequency division multiple access}
\newacronym{fim}{FIM}{Fisher information matrix}

\newacronym{gf}{GFA}{grant free access}
\newacronym{gfdma}{GFDMA}{generalized frequency division multiple access}
\newacronym{ggn}{GGN}{generalized Gaussian noise}
\newacronym{gpqsm}{GPQSM}{generalised pre-coding quadrature spatial modulation}
\newacronym{gps}{GPS}{global positioning system}

\newacronym{g2u}{G2U}{Gaussian to uniform}

\newacronym{hrate}{H-Rate}{high data-rate}
\newacronym{hcas}{HCAS}{hybrid communications and sensing}

\newacronym{ici}{ICI}{inter-channel interference}
\newacronym{ifft}{IFFT}{inverse \gls{fft}}
\newacronym{ipic}{IPIC}{inter-packet interference cancellation}
\newacronym{iui}{IUI}{inter-user interference}
\newacronym{im}{IM}{index modulation}
\newacronym{imma}{IMMA}{index modulation multiple access}
\newacronym{irs}{IRS}{intelligent reflecting surface}
\newacronym{iot}{IoT}{Internet of Things}
\newacronym{iid}{i.i.d.}{independent and identically distributed}
\newacronym{isac}{ISAC}{integrated sensing and communication}

\newacronym{jmld}{JMLD}{joint-multiuser maximum likelihood detector}
\newacronym{jcas}{JCAS}{joint communication and sensing}
\newacronym{kpi}{KPI}{key performance indicator}

\newacronym{led}{LED}{light emitting diode}
\newacronym{llr}{LLR}{log-likelihood ratio}
\newacronym{ldpc}{LDPC}{low-density parity-check}
\newacronym{los}{LoS}{line-of-sight}
\newacronym{Lppm}{$M_n$-PPM}{$M$-ary pulse position modulation}
\newacronym{ls}{LS}{least-square}

\newacronym{mud}{MUD}{multi-user detection}
\newacronym{musa}{MUSA}{multiple-user shared access}
\newacronym{mmse}{MMSE}{minimum mean square error}
\newacronym{mmtc}{mMTC}{massive machine type communication}
\newacronym{manet}{MANET}{mobile ad-hoc network}
\newacronym{mbm}{MBM}{media-based modulation}
\newacronym{mi}{MI}{mutual information}
\newacronym{mimo}{MIMO}{multiple-input multiple-output}
\newacronym{miso}{MISO}{multiple-input single-output}
\newacronym{mrt}{MRT}{maximal ratio transmission}
\newacronym{mld}{MLD}{maximum likelihood detector}
\newacronym{mpsk}{MPSK}{$M_P$-ary phase shift keying}
\newacronym{mqam}{$M_n$-QAM}{$M$-ary quadrature amplitude modulation}
\newacronym{mppm}{MPPM}{multiple pulse position modulation}
\newacronym{mpam}{$M_n$-PAM}{$M$-ary pulse amplitude modulation}
\newacronym{mask}{MASK}{$M_A$-ary amplitude shift keying}
\newacronym{map}{MAP}{maximum a posterior}
\newacronym{mpa}{MPA}{message passing algorithm}
\newacronym{mse}{MSE}{mean squared-error}
\newacronym{mui}{MUI}{multi-user interference}
\newacronym{mmwave}{mmWave}{millimeter wave}
\newacronym{ml}{ML}{machine learning}
\newacronym{mimol}{MIMO}{multiple-input-multiple-output}
\newacronym{mle}{MLE}{maximum likelihood estimation}
\newacronym{music}{MUSIC}{multiple signal cassification}
\newacronym{mrc}{MRC}{maximum ratio combiner}

\newacronym{noma}{NOMA}{non-orthogonal multiple access}
\newacronym{nlos}{nLoS}{non-line-of-sight}

\newacronym{ofdm}{OFDM}{orthogonal frequency division multiplexing}
\newacronym{ook}{OOK}{on-off keying}
\newacronym{oma}{OMA}{orthogonal multiple access}
\newacronym{ofdma}{OFDMA}{orthogonal frequency division multiple access}
\newacronym{onoma}{O-NOMA}{optical-NOMA}

\newacronym{pam}{PAM}{pulse amplitude modulation}
\newacronym{pd}{PD}{power domain}
\newacronym{pep}{PEP}{pairwise error probability}
\newacronym{per}{PER}{packet error rate}
\newacronym{pdf}{PDF}{probability density function}
\newacronym{pa}{PA}{power assignment}
\newacronym{pcb}{PCB}{power coefficient bound}
\newacronym{pmf}{PMF}{probability mass function}
\newacronym{psk}{PSK}{phase shift keying}
\newacronym{psm}{PSM}{precoded spatial modulation}
\newacronym{plnc}{PLNC}{physical layer network coding}
\newacronym{pls}{PLS}{physical layer security}
\newacronym{pascal}{PASCAL}{pilot-aided simultaneous communication and
localization}

\newacronym{qos}{QoS}{quality of service}
\newacronym{qpsk}{QPSK}{quadrature-phase shift keying}
\newacronym{qam}{QAM}{quadrature amplitude modulation}

\newacronym{rf}{RF}{radio frequency}
\newacronym{rhi}{RHI}{residual hardware impairment}
\newacronym{rv}{RV}{random variable}  
\newacronym{ran}{RAN}{radio access network} 
\newacronym{rmse}{RMSE}{root mean square error} 

\newacronym{sc}{SC}{superposition coding}
\newacronym{scfdma}{SC-FDMA}{single-carrier frequency division multiple access}

\newacronym{scma}{SCMA}{sparse code multiple access}
\newacronym{se}{SE}{spectral efficiency}
\newacronym{si}{SI}{self-interference}
\newacronym{ser}{SER}{symbol error rate}
\newacronym{selc}{SeC}{selection combining}
\newacronym{sep}{SEP}{symbol error probability}
\newacronym{sic}{SIC}{successive interference cancellation}
\newacronym{sinr}{SINR}{signal-to-interference-plus-noise ratio}
\newacronym{sm}{SM}{spatial modulation}
\newacronym{snr}{SNR}{signal-to-noise ratio}
\newacronym{spc}{SPC}{short packet communication}
\newacronym{sdma}{SDMA}{space division multiple access}
\newacronym{ssk}{SSK}{space shift keying}
\newacronym{swipt}{SWIPT}{simultaneous wireless information and power transfer}
\newacronym{stlc}{STLC}{space-time line code}
\newacronym{siso}{SISO}{single-input-single-output}
\newacronym{simo}{SIMO}{single-input-multiple-output}
\newacronym{sbs}{sBS}{small BS}
\newacronym{mbs}{mBS}{macro BS}
\newacronym{sca}{SCA}{side channel attacks}

\newacronym{ta}{TA}{transmit amplifier}
\newacronym{tdma}{TDMA}{time division multiple access}
\newacronym{thz}{THz}{terahertz}

\newacronym{ul}{UL}{uplink}
\newacronym{urllc}{URLLC}{ultra-reliable low latency communication}

\newacronym{v2v}{V2V}{vehicle-to-vehicle}
\newacronym{vlc}{VLC}{visible light communication}

\newacronym{wit}{WIT}{wireless information transfer}
\newacronym{wpt}{WPT}{wireless power transfer}
\newacronym{uav}{UAV}{unmanned aerial vehicle}
\newacronym{sat}{SAT}{sinusoidal addition theorem}

\newacronym{op}{OP}{outage probability}
\newacronym{rb}{RB}{resource block}

\newacronym{nr}{NR}{new radio}

 \newacronym{la}{LA}{low altitude}
 \newacronym{ha}{HA}{high altitude}
 \newacronym{leos}{LEOS}{low earth orbit satellite}

\usepackage{float}
\begin{document}
\title{Pilot-Aided Simultaneous Communication And Localization (PASCAL) Under Rician Channels}
\author{Mohammad Al-Jarrah,~\IEEEmembership{Member,
~IEEE}, and Emad Alsusa,~\IEEEmembership{Senior Member,~IEEE,}\vspace{-0.4in}\thanks{M. Al-Jarrah and E. Alsusa are with the Department of Electrical and
Electronic Engineering, University of Manchester, Manchester M13 9PL, U.K.
(e-mail: \{mohammad.al-jarrah, e.alsusa\}@manchester.ac.uk).}\thanks{%
}}


\maketitle

\begin{abstract}
This paper addresses the growing need for modest and reliable localization and communication in wireless networks. We propose a fading-aware Pilot-Aided Simultaneous Communication and Localization (PASCAL) framework that exploits embedded pilot symbols for transmitter localization, in addition to {conventional} channel estimation. A multi-antenna {base station} (BS) is applied, and the propagation channel follows {a} Rician model{,} which takes into account both line-of-sight and non-line-of-sight paths. In addition, to model the temporal correlation between channel realizations that emerges due to {the} transmitter's mobility, {the} Jakes model is adopted, and {a first-order} autoregressive model is used to approximate the errors arising due to outdated channels. A localization algorithm is proposed based on maximum likelihood estimation to obtain optimal device location estimates{,} including the elevation and azimuth angles, the Doppler frequency{,} and distance to the BS. Analytical characterizations, including {the} achievable symbol error rate and the Cramér–Rao lower bound, which is adopted as a benchmark for the localization errors, are derived to quantify the communication performance and the localization limits. Theoretical analysis corroborated by simulation results demonstrates that PASCAL operates near fundamental localization performance limits under certain conditions{,} such as high signal-to-noise ratio, and in general, it performs better than the widely used {Multiple Signal Classification} (MUSIC) algorithm{,} while sustaining reliable communication across several operating conditions.

\end{abstract}
\glsresetall
\section{Introduction}

After the successful deployment of the fifth-generation (5G) networks worldwide, many technologies, services, and applications have emerged. Autonomous and self-driving vehicles, \glspl{uav}, mobile indoor and outdoor robotic platforms, and the ever-expanding domain of the \gls{iot} now rely on wireless networks; not only for high-rate data transfer but also for precise and continuous awareness of their surrounding environment. These developments have placed unprecedented demands on modern wireless systems, particularly in terms of providing accurate, robust, and real-time localization services alongside the traditional connectivity functions. As research efforts advance toward the sixth generation (6G) of wireless networks, the scope of network responsibilities is expected to broaden significantly. Beyond conventional communication, 6G systems are anticipated to support a wide range of integrated sensing capabilities, including detection, localization, tracking, navigation, and environmental surveillance. Such \gls{isac} services are essential for enabling reliable autonomous operation, improving situational awareness, and supporting safety-critical applications in complex and dynamic environments such as dense urban areas, industrial facilities, and transportation systems \cite{10274474, 7849233, 10113891}. However, these expanded functionalities introduce additional challenges. Wireless networks operate with inherently limited spectrum, power, and computational resources, and the addition of high-precision sensing tasks intensifies the pressure on these constraints. Designing lightweight algorithms capable of performing fast, accurate detection, sensing, and localization while meeting real-time latency requirements becomes significantly more demanding.
\subsection{Literature survey}
In recent years, extensive research toward unifying communication and sensing on the same platform has been conducted by academia and industry. The concept
of \gls{isac} is introduced in \cite{8386661, 8288677}, in which a \gls{bs}
is applied to provide dual communication-sensing operation. Two models have been introduced which are referred to separated and
shared deployments. In the separated deployment, the \gls{bs} antennas are
distributed among each functionality, whereas the whole antenna array is exploited for both in the
shared deployment scenario. Several designs for the signal waveforms and
beampatterns are proposed in \cite{8386661, 8288677, yousef_arch} to satisfy the requirements of communication users’ rates and detection capability
of the radar sub-system. Moreover, various signal
processing tools that can be applied for \gls{isac} systems can
be found in \cite{9540344} for three possible scenarios, radar-centric, communication-centric, and joint design. Furthermore, resource allocation algorithms have been presented in the literature, where power allocation and transmission strategies are
proposed with the aim of optimizing the performance of coexisting radar and communication systems \cite{9359480, 8834831, 10507154}. Obviously, there is a persistent need for including sensing services in the cellular network architecture. However, \gls{isac} schemes require careful integrated signal design, which usually ends up involving complex optimization methods and increases the computational complexity burden on \glspl{bs}, which impacts real-time applications. Additionally, the allocation of energy and spectrum resources for sensing, target detection, and localization can degrade the communication performance, as \glspl{bs} are typically constrained by limited resources.

A prominent research area, within the broader concept of sensing within telecommunication networks, leverages the fact that devices such as \glspl{uav}, smart vehicles, and mobile users are active nodes capable of both transmitting and receiving signals, allowing their communication signals to be utilized for sensing and localization tasks. Accordingly, by exploiting their ability to transmit a known signal periodically and/or the data frame structure, which typically contains pilot signals, these devices can be localized in a very efficient way, achieving simultaneous localization and communication. For instance, in \cite{11, 12}, a pilot signal is inserted in the transmitted data sequence sent by the \gls{bs} and then the data sequence is sent to the receiving device. The \gls{bs} is also equipped with a radar unit that relies on the reflected echo of these pilots for sensing the channel and localizing targets. Driven by the fact that pilot signals can be very beneficial for sensing and localization, the design of an optimal data frame structure with pilots is also investigated in \cite{13,14}. In \cite{15}, the design of a novel symbol placement scheme to facilitate channel estimation and localization is presented for orthogonal time-frequency-space transmission, in which the pilot and
data symbols are superimposed. In \cite{16}, a joint pilot optimization for a target
detection and channel estimation scheme is proposed assuming that the radar and communication channels exhibit a certain joint burst sparsity. However, the authors have considered target detection only whereas localization has not been investigated. In \cite{23}, a system model consists of a \gls{uav} acting as a \gls{bs} and equipped with a rectangular array, and transmitting a sequence of pilot and data symbols to a ground node is introduced, where the \gls{uav} is responsible for delivering data to the ground node and estimating its channel and location using the echo of the transmitted pilots. However, a \gls{los} channel is assumed, which limits the applicability of the introduced algorithms since channels usually suffer from \gls{nlos} propagation which impacts the \gls{los} received signal. In \cite{24}, a \gls{uav} that gathers data from a number of ground nodes is deployed, where uploaded data from these ground nodes are exploited to estimate the its trajectory. The estimated trajectory is then exploited to optimize the UAV trajectory to enhance the communication links. \vspace{-.1in}

\subsection{Motivation and contribution}
Apparently, the demand for effective dual communication-localization capabilities in modern wireless systems is rapidly increasing, driven by the need to deliver high data rates while simultaneously enabling precise sensing and positioning within a unified platform. However, typical \gls{isac} architectures require the \gls{bs} to allocate a portion of its communication resources for sensing, inherently creating a tradeoff between communication efficiency and sensing or localization accuracy. These systems also rely primarily on echo-based localization, similar to radar perception, which incurs substantial power loss over the two-way propagation path, i.e., the \gls{bs}-target and target-\gls{bs} path-loss. Furthermore, civilian \glspl{uav} such as drones and quadcopters often have very small radar cross-section area, resulting in low reflection coefficients, thereby making reliable localization extremely challenging under conventional echo-based approaches.

On the other hand, although research efforts targeting the localization of active devices in \cite{13,14,15,16,23,24} address some of the limitations discussed above, several challenges remain. These approaches continue to rely on target-reflected signals, which are vulnerable to substantial clutter and exhibit low \gls{snr} as a consequence of two-way path loss and the small radar cross-section of the active device. To overcome these limitations, our previous work \cite{shuai1, shuai2} have introduced \gls{pascal}, in which pilot symbols embedded within uplink data frames at fixed intervals are exploited for both channel estimation and localization. In \gls{pascal}, the \gls{bs} leverages these uplink frames to estimate the channel, localize the transmitting nodes using the embedded pilots, and subsequently use the estimated channel for data detection. Unlike our earlier studies, which were limited to the \gls{los} channel assumption, the present work investigates PASCAL under a more realistic Rician fading model, incorporating the Clarke–Jakes formulation to capture the temporal correlation of the time-varying channel caused by mobility. Table \ref{tab1} summarizes the key advantages offered by \gls{pascal} relative to conventional \gls{isac} frameworks. To this end, the contributions of this paper are summarized as follows:

\begin{enumerate}

    \item We develop practical system and channel models for PASCAL operating under Rician fading, capturing both \gls{los} and \gls{nlos} components and incorporating temporal correlation arising from mobility through the Clarke-Jakes model.

    \item We propose a robust estimator for the transmitter parameters, including the azimuth and elevation angles, the Doppler frequency and the distance, with the aid of the \gls{mle} framework. Closed-form optimal expressions are derived for the estimates of Doppler frequency and distance given the other parameters, whereas optimization of \gls{mle} is used for estimating the azimuth and elevation angles. The estimation of all parameters is carried out alternatively and iteratively to obtain the final estimates of all parameters.
    
    \item An analytical characterization for the achievable performance of the considered system model. The \gls{crlb} is derived to provide a benchmark for the limits of the performance of elevation and azimuth angles estimation, whereas the estimation errors in the distance and Doppler frequency estimates are found in closed forms. In addition, for the sake of completeness, the achievable \gls{ser} is derived under the Rician channel, channel estimation errors and the temporal correlation arising due to mobility.
    
    \item We perform comprehensive Monte Carlo simulations to validate the analytical results and assess PASCAL in various operating conditions. The obtained results show that the framework closely approaches theoretical localization limits while maintaining robust communication under several configurations.

\end{enumerate}
\color{black}
\begin{table*}[ht] 
\centering
\renewcommand{\arraystretch}{1.4} 
\setlength{\tabcolsep}{10pt}       
\caption{Comparison between \gls{pascal} and conventional \vspace{-.06in}\gls{isac}}
\begin{tabularx}{\textwidth}{|c|X|X|}
\hline
\rowcolor{gray!20}
\textbf{Aspect} & \textbf{Pilot-Aided Localization} & \textbf{Conventional \gls{isac}} \\ \hline
Waveform design & Uses known pilot symbols or pilot sequences & Requires robust joint waveform design \\ \hline
Synchronization & High precision due to known pilot timings & Depends on joint waveform and channel estimation accuracy \\ \hline
Estimation accuracy & Usually higher since pilots are known and signal goes through a single way & Lower depending on waveform constraints. \\ \hline
Spectral efficiency & Higher (pilots already embedded) & Lower (requires dedicated sensing waveform) \\ \hline
Complexity & Simpler estimation algorithms & Joint processing is needed for communication and sensing parameters \\ \hline
Use case & Active connected devices & Passive and non-cooperative targets \\ \hline
\end{tabularx}\label{tab1}\vspace{-0.2in}
\end{table*}

The remainder of this paper is organized as follows. Section II details the system and channel models. Section III provides algorithms for estimation and localization and their performance analysis. Section IV investigates the communication aspect including the analysis of \gls{ser}. Simulation results are provided in Section V, and Section VI concludes the paper.\vspace{-.1in}

\section{System model}
This work considers a \gls{mimo} \gls{bs} equipped with an $N_{%
\mathrm{x}} \times N_{\mathrm{y}}$ uniform rectangular array, with $N_{\mathrm{x}}$
and $N_{\mathrm{y}}$, respectively, denoting the number of antennas in the $x$ and 
$y$ directions, respectively. The \gls{bs} is deployed to receive data symbols from $K$ single-antenna \glspl{cv}. Narrowband transmission with $\mathcal{M}$-ary \gls{psk} signaling is considered for transmitting data symbols to the \gls{bs} in the uplink
direction with single-carrier modulation. In addition, orthogonal signaling,
such as  \gls{tdma} or \gls{fdma}, is adopted to serve multiple \glspl{cv}. Moreover, similar to traditional communication systems, a pilot
signal with a known transmitted symbol is inserted at the beginning of each
group of $T$ data symbols, which is utilized to estimate the channel gain
which experiences a single-tap Rician fading\footnote{A single-tap channel is appropriate for the considered narrowband communication.}, where multiple reflections from the environment are received in addition to the direct \gls{los} signal component. The pilot signal is also exploited in our work to estimate the mobility and location
parameters of the \glspl{cv}, including the azimuth and elevation angles, the Doppler
frequency that quantifies the velocity, and the distance to the \gls{bs}. Since \glspl{cv}
employ orthogonal multiple access, we focus here on the analysis of a single \gls{cv} while the
same procedures can be applied to all other \glspl{cv}.\vspace{-.1in}

\subsection{Received signal model}

At the $l$th signaling period, the \gls{cv} draws a data symbol from a \gls{psk} constellation\footnote{The algorithms introduced in this paper are applicable for any modulation type but the derived \gls{ser} considers PSK.}, $s_{i,\mathrm{d}}\left[ l\right] $, or a pilot signal $s_{i,%
\mathrm{p}}\left[l_\mathrm{p}\right] $, which is sent once every $T$ time slots. Then, the \gls{cv} sends the selected symbol to the \gls{bs}, where the received signal
vector at the \gls{bs} can be written as follows
\begin{equation}
\mathbf{y}_{i,\mathrm{d|p}}\left[ l\right] =\sqrt{p_{i,\mathrm{d|p}}}\mathbf{%
h}_{i}\left[ l\right] s_{i,\mathrm{d|p}}\left[ l\right] +\mathbf{n}_{i}\left[
l\right]  \label{E-y-1}
\end{equation}%
where $p_{i,\mathrm{d}}$ is the transmit power for the $i$th CV dedicated to
transmitting data symbols, $p_{i,\mathrm{p}}$ is the transmit power dedicated
to the pilot signal that is typically set to $s_{i,\mathrm{p}}\left[ l%
\right] =1\mathrm{e}^{j0}$, $\mathbf{h}_{i}\left[ l\right] \in \mathbb{C}%
^{1\times \mathcal{N}}$ with $\mathcal{N}=N_{\mathrm{x}} N_{\mathrm{y}%
} $ is a vector of the channel coefficients, and $\mathbf{n}_{i}\left[ l%
\right] \sim \mathcal{CN}\left( 0,\sigma _{n}^{2}\mathbf{I}_{\mathcal{N}%
}\right) $ is the \gls{awgn}. 

Similar to typical communication systems, pilots, also known as reference
signals because they are typically known to the receiver, are transmitted at specific time slots.
Without loss of generality, one pilot signal is sent at the beginning of a
frame consisting of $T$ symbols. Upon receiving the pilots, the receiver can
infer the channel, which is applied for detecting the subsequent data symbols in
the next time slots until the next pilot is received. Accordingly, let us define $\forall m \in \{1,2,\ldots ,M\},\forall t \in
\{1,2,\ldots ,T \}$ to indicate  the frame number and symbol order in the $m$th frame, respectively, for example, $t=1$ is the position of the pilot $\forall m$. Thus, by referring to the time index $l$ in \eqref{E-y-1}, we obtain $t=((l-1)\;\mathrm{mod}\; T)+1$ and $m=\frac{l-t}{T}+1$, or equivalently, $l=(m-1)T+t$. We will use the indices $[m,t]$ instead of $l$ henceforth to connect the analysis to the frame and symbol indices. It is worth noting that this frame structure, i.e., a pilot followed by $T-1$ data symbols, is standard in wireless communications. The main objective of our work is to add localization functionality to the system by utilizing the same frame structure to locate the transmitting \gls{uav} by estimating its angular position, Doppler frequency, and distance.\vspace{-0.2in}

\subsection{Channel model}

The propagation medium from UAV to BS is subject to distance-dependent large-scale path-loss and small-scale fading as detailed in the discussion below.

\subsubsection{Path-loss model}

In general, the large-scale path loss is typically dependent on the
distance between the transmitter and receiver, which makes the received
signal strength informative about the range, i.e., the distance to the transmitter.
The amount of path loss can be expressed as \cite{rapp, PMLE, rfid}%
\begin{equation}
B_{i}\left( d_{i},f_{c}\right) =G_{\mathrm{T}}G_{\mathrm{R}}\left( \frac{c}{%
4\pi f_{c}}\right)^2 \left( \frac{d_{0}}{d_{i}}\right) ^{\kappa }  \label{e-pl}
\end{equation}%
where $d_{i}$ is the distance between vehicle $i$ and the BS, $d_{0}$ is a
reference distance, which is typically set to $1$ m, $\kappa $ is defined as
the path-loss exponent with values ranging from $2$ to $4$, $c=3 \times 10^{8}$ m/s is the speed
of light, and $f_{c}$ is the operating frequency. Without loss of
generality, we assume that the antenna gains are normalized to unity, i.e., $G_{%
\mathrm{R}}=G_{\mathrm{T}}=1$. Given the vector of the actual location of the $i$th \gls{cv}
in Cartesian coordinates, $\mathbf{p}_{i}=[x_{i},y_{i},z_{i}]\in \mathbb{R}%
^{1\times 3}$, and the location of the BS, which is known in
advance, $\mathbf{p}_{\mathrm{BS}}=\left[ x_{%
\mathrm{BS}},y_{\mathrm{BS}},z_{\mathrm{BS}}\right] $, the actual distance between CV $i$ and the BS is%
\begin{eqnarray}
d_{i} &\triangleq &\left\Vert \mathbf{p}_{i}-\mathbf{p}_{\mathrm{BS}%
}\right\Vert \\
&=&\sqrt{\left( x_{i}-x_{\mathrm{BS}}\right) ^{2}+\left( y_{i}-y_{\mathrm{BS}%
}\right) ^{2}+\left( z_{i}-z_{\mathrm{BS}}\right) ^{2}}
\end{eqnarray}

\subsubsection{Small-scale Rician fading with mobility}\label{sec-rician-model}
The small-scale fading for CV-BS links obeys the Rician model in which the
received signal has an \gls{los} component in addition to the \gls{nlos} component that
arises due to multipath reflections. The \gls{nlos} component randomly affects
the envelope and phase of the received signal. According to the Rician model, as the Rician factor $%
\mathcal{K}$ increases,
the \gls{los} component becomes more dominant; for example, the best link quality is achieved when $%
\mathcal{K\!\rightarrow \!\infty }$ which is equivalent to ideal \gls{los}
propagation. Let us denote the channel coefficient from the $i$th CV to the $%
\dot n$th antenna of the BS, which is an element of $\mathbf{h}_{i}\left[ m,t\right]
$, by $h_{i,\dot n}\left[ m,t\right] $ $\forall \dot{n}\in \{1,2,\ldots ,\mathcal{%
N}\}$, i.e., $\mathbf{h}_{i}\left[ m,t\right] =\left[ h_{i,1}\left[ m,t%
\right] ,h_{i,2}\left[ m,t\right] ,\ldots ,h_{i,\mathcal{N}}\left[ m,t\right]
\right] $. The subscripts $m$ and $n$ here denote the frame and antenna indices, respectively, which replace the time index $l$, and thus $\mathbf{h}_i[m,t]$ is the same channel vector $\mathbf{h}[l]$ but with different indexing that is applied to make the presentation more concise and readable. By incorporating the path-loss model, in a single-tap
channel where delays from all paths are received at the same time instant, $%
h_{i}\left[ m,t\right] $ can be generally expressed as \cite%
{8787874,irs} 
\begin{multline}
\!h_{i,\dot{n}}\!\left[ m,t\right] \!=\!\left( \sqrt{\!B_{i}\left( d_{i},f_{c}\right) }
\mathbf{a}_{\dot{n}}\!\left(\! \theta _{i}\left[ m,t\right] ,\phi _{i}\!\left[ m,t%
\right] \right) h_{i,\mathrm{LoS}}\left[ m,t\right] \right.\\
+\left. h_{i,\dot{n},\mathrm{%
nLoS}}\left[ m,t\right] \right)
\end{multline}
where $\mathbf{a}_{\dot{n}}\left( \theta _{i}\left[ m,t\right] ,\phi _{i}%
\left[ m,t\right] \right) $ is the $\dot{n}$th element of the array steering
vector $\mathbf{a}\left( \theta _{i}\left[ m,t\right] ,\phi _{i}\left[ m,t%
\right] \right) \in \mathbb{C}^{\mathcal{N}\times 1}$ which is a function of the elevation and azimuth angles of the CV. The \gls{los} component $h_{i,\mathrm{LoS}}$ is a function of the Doppler frequency, i.e., $h_{i,\mathrm{LoS}}\left[ m,t%
\right] =\mathrm{e}^{-j2\pi f_{D,i}(l-1)T_{s}}$ with $l=(m-1)T+t$, $f_{D,i}$ is the maximum
Doppler frequency that is a function of the velocity of the $i$th CV, and $T_{s}$
is the symbol duration in seconds\footnote{It should be noted that the Doppler effect causes a time-varying phase rotation that is proportional to the elapsed time. Thus, at the reference instant $l=0$, no time has passed, so there is no accumulated phase, and thus we subtract $1$ from $l$ in the exponent.}. The \gls{nlos} term, $h_{i,\dot{n},\mathrm{%
nLoS}}$, which emerges due to a large number of scatterers in
the environment, is typically modeled as a complex normal random variable, and thus $h_{i,\dot{n},\mathrm{nLoS}}%
\left[ m,t\right] $ is a Rayleigh fading term, i.e., $h_{i,\dot{n},\mathrm{nLoS}}%
\left[ m,t\right] \sim \mathcal{CN}\left( 0,\sigma _{h}^{2}\right) $. It is
noteworthy observing that the Rician factor is the power ratio between \gls{los}
and \gls{nlos} signal components, that is given by $\mathcal{K=}\frac{{\!B_{i}\left( d_{i},f_{c}\right) }\left\vert \mathbf{a}%
_{\dot{n}}\left( \theta _{i}\left[ m,t\right] ,\phi _{i}\left[ m,t\right]
\right) h_{i,\mathrm{LoS}}\left[ m,t\right] \right\vert ^{2}}{\sigma _{h}^{2}%
}=\frac{{\!B_{i}\left( d_{i},f_{c}\right) }}{%
\sigma _{h}^{2}}$. 

 Obviously, the localization is challenging due to the
variations in the channel, as well as the existence of multipath rays which
do not contribute to the actual location of the CV since they are reflected from
other objects in the environment. The $(m,n)$th antenna array element's response can be given by\vspace{-.05in}
\begin{multline}
a_{m,n}(\phi_i,\theta_i)
= \exp\!\left(
-j\frac{2\pi \Delta}{\lambda}
\Big[(m-1)\cos\phi_i\sin\theta_i \right.+ (n-1)\\
\left.\times \sin\phi_i\sin\theta_i\Big]
\right),
\quad
m=1,\dots,N_\mathrm{x},\;
n=1,\dots,N_\mathrm{y}.\label{e-array-resp-1}
\end{multline}
with a vectorized version represented by\vspace{-.05in}
\begin{equation}
\mathbf{a}(\phi_i,\theta_i)
=
\begin{bmatrix}
a_{1,1}(\phi_i,\theta_i),
a_{1,2}(\phi_i,\theta_i),
\cdots,
a_{M,N}(\phi,\theta)
\end{bmatrix}^T\label{e-array-resp-2}
\end{equation}
where $\Delta$ is the spacing between adjacent array elements.
\subsubsection{Autoregressive model\label{sec-autoreg}}

Due to the mobility of the CV, channels estimated at a certain time instant
using pilots will become outdated when the data symbols are received which may
result in erroneous data detection at high speeds. To model the outdated channel effect, we leverage the first-order autoregressive model
to characterize the outdated \gls{nlos} channel component, and thus $\mathbf{h}_{i}%
\left[ l\right] $ can be represented as
\begin{equation}
\mathbf{h}_{i}\left[ l\right] =\mathbf{\tilde{h}}_{i,\mathrm{LoS}}\left[ l\right] +%
\mathbf{h}_{i,\mathrm{nLoS}}\left[ l\right] 
\end{equation}
which can be expressed as
\begin{multline}
\mathbf{h}_{i}\left[ l\right] = \sqrt{B_{i}\left( d_{i}\left[ l\right]
,f_{c}\right) } \mathbf{a}\left( \theta _{i}\left[ l\right] ,\phi _{i}%
\left[ l\right] \right) \mathrm{e}^{-j2\pi f_{D,i}(l-1)T_{s}}   \\
+  \rho \left(
l\right) \mathbf{h}_{i,\mathrm{nLoS}}\left[ l_\mathrm{p}\right] +\sqrt{1-\rho
\left( l\right) ^{2}}\mathbf{\ddot{n}}_{i,\mathrm{nLoS}}\left[ l\right]
   \label{E-ARM-with-Los}
\end{multline}
where $\mathbf{\tilde{h}}_{i,\mathrm{LoS}}[l]\!=\!\sqrt{\!B_{i}\!\left( d_{i}\left[ l\right]
\!,f_{c}\right) } \mathbf{a}\!\left( \theta _{i}\!\left[ l\right]\! ,\!\phi _{i}\!%
\left[ l\right] \right) \mathrm{e}^{-j2\pi f_{D,i}(l-1)T_{s}}$ is the effective or equivalent \gls{los} channel vector, and $\mathbf{\ddot{n}}_{i}\left[ l%
\right] \sim \mathcal{CN}\left( 0,\sigma _{\ddot{n}}^{2}\mathbf{I}_{\mathcal{N}%
}\right) $ is a Gaussian random process.
It should be noted that for typical values of CV speeds and symbol rates, the change in $%
d_{i}\left[ l\right]$ during a frame can be neglected. According to Jakes model, the relative mobility of the transmitter and receiver, where the scatterers are stationary, induces a temporal correlation between channel samples with a correlation
coefficient between two time instants $l$ and $\acute{l}$ given by $\rho
_{i}( l-\acute{l}) =\sigma _{h}^{2}J_{0}\left( 2\pi f_{D,i}(
l-\acute{l}) T_{s}\right) $ with $J_{0}\left( \cdot \right) $
denoting the Bessel function of the first kind and $0$th order.\vspace{-.1in}

\section{Channel Estimation and
Localization}

In this section, we develop algorithms for the effective implementation of PASCAL by relying on the pilot signals for the estimation of the CV mobility parameters, i.e., $\theta_i$, $\phi_i$, $f_{D,i}$, and $d_i$, and the detection of data symbols. We also provide a review of pilot-based channel estimation for the sake of completeness. Moreover, according to Jakes model, the relative mobility of the transmitter and receiver, where the scatterers are stationary, induces a temporal correlation between channel samples with a correlation
coefficient between two time instants $l$ and $\acute{l}$ given by $\rho
_{i}( l-\acute{l}) =\sigma _{h}^{2}J_{0}\left( 2\pi f_{D,i}(
l-\acute{l}) T_{s}\right) $ with $J_{0}\left( \cdot \right) $
denoting the Bessel function of the first kind and $0$th order, and thus the
\gls{nlos} components can be informative of Doppler frequency according to this
model\color{black}. This is an interesting observation which can enhance the Doppler frequency estimation using \gls{nlos} scattered components.\vspace{-0.2in}

\subsection{Channel estimation}

Based on the received pilot signals $\mathbf{y}_{i,\mathrm{p}}\left[ m,t=1%
\right] \forall m$ in (\ref{E-y-1}), the channel estimator can be optimally derived using \gls{mle}. After some simplification, the channel estimate can be written as
\begin{equation}
\mathbf{\hat{h}}_{i}\left[ m,1\right] =\frac{1}{\sqrt{p_{i,\mathrm{p}}}}%
\mathbf{y}_{i,\mathrm{p}}\left[ m,1\right]\label{e-est-ch-1}
\end{equation}%
It can be shown that the mean of the estimator is unbiased since $\mathrm{E}\left[ \mathbf{\hat{h}}_{i}\left[ m,1%
\right] \right] =\mathbf{h}_{i}\left[ m,1\right] $. The \gls{mse} of this estimator is%
\begin{equation}
\mathrm{MSE}_{i}\left[ m,1\right] =\frac{1}{\mathcal{N}}\mathrm{E}\left[
\left\Vert \mathbf{\hat{h}}_{i}\left[ m,1\right] -\mathbf{h}_{i}\left[ m,1%
\right] \right\Vert ^{2}\right] =\frac{\sigma _{n}^{2}}{p_{i,\mathrm{p}}} \label{e-mse-ch-1}
\end{equation}
where $\left\Vert \mathbf{\cdot }\right\Vert ^{2}$ is the squared Euclidean norm. Nevertheless, it should be noted that the estimation error statistics shown
in \eqref{e-mse-ch-1} {represent} the errors {resulting from} using the pilot to estimate the channel at the
instant of the pilot signalling period. However, this channel estimate
is used to decode the subsequent data symbols, for which the channel becomes outdated due to CV's mobility. Therefore, by taking this into account, the equivalent error at the $%
t$th signalling slot in the $m$th frame for all subsequent time slots $t=\left\{
2,3,\dots,T\right\} $ is a function of the CV's parameters, i.e., $\{\theta _{i},\phi _{i},f_{D,i},d_{i}\}$, and can be derived as
\begin{equation}
\mathrm{MSE}_{i}\left[ m,t|\theta _{i},\phi _{i},f_{D,i},d_{i}\right] =
\frac{1}{\mathcal{N}}\mathrm{E}\left[ \left\Vert \mathbf{\hat{h}}_{i}\left[ m,t%
\right] -\mathbf{h}_{i}\left[ m,t\right] \right\Vert ^{2}\right] 
\end{equation}
$\forall t=\left\{
2,3,\dots,T\right\} $. Since the channel is estimated using the inserted pilot signal at $t=1$, and by noting that $\mathbf{\hat{h}}_{i}\left[ m,t=1
\right]=\mathbf{{h}}_{i}\left[ m,t=1%
\right] +\mathbf{\hat{%
h}}_{i}^{\mathrm{err}}\left[ m,t=1\right]$ with $\mathbf{\hat{%
h}}_{i}^{\mathrm{err}}\left[ m,t=1\right]$ denoting the estimation error of $\mathbf{{%
h}}_{i}\left[ m,t=1\right]$, we obtain
\begin{multline}
    \mathrm{MSE}_{i}\left[ m,t|\theta _{i},\phi _{i},f_{D,i},d_{i}\right] =
\frac{1}{\mathcal{N}}\\
 \times\mathrm{E}\left[ \left\Vert \mathbf{{h}}_{i}\left[ m,t=1%
\right] +\mathbf{\hat{%
h}}_{i}^{\mathrm{err}}\left[ m,t=1\right]-\mathbf{h}_{i}\left[ m,t\right] \right\Vert ^{2}\right]\label{e-mse-13}
\end{multline}
$\forall t=\left\{
2,3,\dots,T\right\} $. 

By using the autoregressive model in \eqref{E-ARM-with-Los} and noting that $\mathrm{E}\left[\mathbf{\hat{%
h}}_{i}^{\mathrm{err}}\left[ m,t=1\right]\right]=\mathrm{E}\left[ \mathbf{h}_{i,\mathrm{nLoS}}\left[l\right] \right]=\mathrm{E}\left[ \mathbf{\ddot{n}}_{i,\mathrm{nLoS}}\left[ l\right]\right]=0$, as well as with the aid of the identity $\mathbb{E}\!\left[\|\mathbf{a} + \mathbf{b} +\mathbf{c}\|^2\right] 
= \|\mathbf{a}\|^2 + \mathbb{E}\!\left[\|\mathbf{b}\|^2\right] + \mathbb{E}\!\left[\|\mathbf{c}\|^2\right]$ for a deterministic $\mathbf{a}$ and zero-mean $\mathbf{b}$ and $\mathbf{c}$, the \gls{mse} in \eqref{e-mse-13} can be derived as given in \eqref{e-long-1} on the top of page \pageref{e-long-1}, where $\sigma _{\mathrm{err}}^{2}$ denotes the variance of the estimation error in \eqref{e-est-ch-1}, which is given in \eqref{e-mse-ch-1}, where $l_\mathrm{p}=(m-1)T+1$ is the pilot index, and $l=(m-1)T+t$ is the data symbol index.

\begin{figure*}[tbp]

\begin{eqnarray}
\mathrm{MSE}_{i}\!\!\left[ m,t|\theta _{i},\phi _{i},f_{D,i},d_{i}\right]\!\!&\!\!=\!&\!\!\frac{1}{\mathcal{N}}\!\!\left( \left\Vert \sqrt{B_{i}\!\left( d_{i}\left[ l_\mathrm{p}%
\right] ,f_{c}\right) }\mathbf{a}\left( \theta _{i}\left[ l_\mathrm{p}\right] ,\phi
_{i}\left[ l_\mathrm{p}\right] \right)\mathrm{e}^{-j2\pi f_{D,i}(l_\mathrm{p}-1)T_{s}}\! -\!\sqrt{B_{i}\!\left( d_{i}\left[ l\right]
,f_{c}\right) }\mathbf{a}\left( \theta _{i}\left[ l\right] ,\phi _{i}\left[ l%
\right] \right) \right. \right. \nonumber \\
&&\left.\left. \times \mathrm{e}^{-j2\pi f_{D,i}(l-1)T_{s}}\right\Vert ^{2} +\mathcal{N}%
\sigma _{\mathrm{err}}^{2} +\sigma _{h}^{2}\mathcal{N}(1-\rho(l))^2+ \left( 1-\rho \left( l\right) ^{2}\right) \mathcal{N}\sigma _{\mathbf{%
\ddot{n}}}^{2}\right)\label{e-long-1}
\end{eqnarray}\vspace{-0.1in}
\end{figure*}
The applied outdated channel estimates for equalization $\forall t=\left\{
2,3,\dots,T\right\} $ are biased and the amount of bias is given by $\mathrm{B}_{i}\left[ l|\theta _{i},\phi _{i},f_{D,i},d_{i}\right] =\frac{1%
}{\mathcal{N}}\mathrm{E}\left[ \mathbf{\hat{h}}_{i}\left[ l\right] -\mathbf{h%
}_{i}\left[ l\right] \right]$ which can be derived as
\begin{multline}
\mathrm{B}_{i}\left[ l|\theta _{i},\phi _{i},f_{D,i},d_{i}\right] 
\!=\\
\!\frac{1}{\mathcal{N}}\!\!\left( \sqrt{\!B_{i}\left( d_{i}\left[ l_\mathrm{p}\right]
,f_{c}\right) }\mathbf{a}\!\left( \theta _{i}\left[ l_\mathrm{p}\right] ,\phi _{i}%
\left[ l_\mathrm{p}\right] \right) \mathrm{e}^{-j2\pi f_{D,i}(l_\mathrm{p}-1)T_{s}}\right.
\\-\left.\sqrt{B_{i}\left( d_{i}\left[ l\right]
,f_{c}\right) }\mathbf{a}\left( \theta _{i}\left[ l\right] ,\phi _{i}\left[ l%
\right] \right) \mathrm{e}^{-j2\pi f_{D,i}(l-1)T_{s}}\right)
\end{multline}

\subsection{Mobility Parameters estimation using MLE}
Thereafter, it can be noted that the estimated channel vector obtained from the pilot signal at the time instance $[m,t=1]$ contains
information about the CV's mobility and location parameters, and thus these parameters can be extracted from the estimated channels. Based on the channel model in Sec. \ref{sec-rician-model}, $\mathbf{\hat{h}}_{i}\!\left[ m,t=1\right]\!$ can be expressed as 
\begin{multline}
\!\!\mathbf{\hat{h}}_{i}\!\left[ m,t=1\right]\! =\! \sqrt{\!B_{i}\left(
d_{i},f_{c}\right) }\mathbf{a}\!\left( \theta _{i}\left[  m\!,t=1\right]
,\phi _{i}\left[  m,t=1\right] \right) \\
 \times\mathrm{e}^{-j2\pi f_{D,i}(l_\mathrm{p}-1)T_{s}}+%
\mathbf{h}_{i,\mathrm{nLoS}}\left[  m,t=1\right] +\frac{\mathbf{n}_i}{\sqrt{p_{i,\mathrm{p}}}}
\label{e-h-hat}
\end{multline}
Interestingly, as can be observed from (\ref{e-h-hat}), $\mathbf{h}_{i,\dot{n%
},\mathrm{nLoS}}\left[ m,t=1\right] $ acts as additive noise with an effective \gls{snr}
represented by the Rician factor $\mathcal{R=}\frac{{p_{i,\mathrm{p}}}{\!B_{i}\left(
d_{i},f_{c}\right) }\left\vert h_{i,\mathrm{%
LoS}}\left[ m,t=1\right] \right\vert ^{2}}{{p_{i,\mathrm{p}}}\sigma _{ h}^{2}+\sigma_n^2}$.

In the following, we establish an algorithm to extract the mobility parameters using the \gls{mle} theorem.  The elements of $\mathbf{h}_{i,\mathrm{nLoS}}\left[  m,t=1\right]$  are {spatially} \gls{iid} Gaussian random variables, each with a mean of $0$
and a variance of $\sigma _{h}^{2}$. Therefore, by considering (\ref{e-h-hat}%
), the MLE problem for $\Psi[l_\mathrm{p}]\triangleq\left[ \theta _{i}\left[l_\mathrm{p}\right] ,\phi _{i}\left[l_\mathrm{p}\right]
,f_{D,i}\left[l_\mathrm{p}\right] \right] $\ can be formulated as 
\begin{multline}
\!\!\hat \Psi[l_\mathrm{p}]\!=\!\underset{\Psi[l_\mathrm{p}] }{%
\arg \min}\!\left\Vert \mathbf{\hat{h}}_{i}\!\left[l_\mathrm{p}\right] \!-\!
\sqrt{\!\!B_{i}\!\left( d_{i}\left[l_\mathrm{p}\right] \!,\!f_{c}\right) }\mathbf{a}\!\left(
\theta _{i}\left[l_\mathrm{p}\right] \!,\!\phi _{i}\left[l_\mathrm{p}\right] \right) \right.\\
\left.\times \mathrm{e}^{-j2\pi f_{D,i}(l_\mathrm{p}-1)T_{s}}\!\right\Vert
^{2}  \label{e-mleee}
\end{multline}
Let us define $    \mathcal J(\Psi)
    =
    \left\|
        \hat{\mathbf h}[l_p]
        -
       { \ddot {\mathbf {s}}}(\Psi)
    \right\|^2$ with ${ \ddot {\mathbf {s}}}(\Psi)=\sqrt{B(d_i[l_p],f_c)}\,\mathbf{a}(\theta_i[l_p],\phi_i[l_p])\,
    e^{-j2\pi f_{D.i} (l_p-1)T_s}$. Since $\mathcal J(\Psi)$ is real-valued and depends on real-valued
parameters, its gradient is given by
\begin{equation}
\label{eq:real_gradient}
    \nabla_{\Psi}\mathcal J (\Psi)
    =
    -2\,\Re\!\left\{
        \mathbf J_{\ddot{\mathbf s} (\Psi)}^H
        \left(
            \hat{\mathbf h}[l_\mathrm{p}]
            -
            \ddot{\mathbf s} (\Psi)
        \right)
    \right\},
\end{equation}
where $\mathbf J_{\ddot{\mathbf s} (\Psi)}$ is the Jacobian of $\ddot{\mathbf s}$, i.e., $\mathbf J_{\ddot{\mathbf s} (\Psi)}
    =
    \frac{\partial \ddot{\mathbf s} (\Psi)}{\partial \Psi}
    =
    \begin{bmatrix}
        \frac{\partial \ddot{\mathbf s} (\Psi)}{\partial d_i[l_p]} &
        \frac{\partial \ddot{\mathbf s} (\Psi)}{\partial f_{D,i}} &
        \frac{\partial \ddot{\mathbf s} (\Psi)}{\partial \theta_i[l_p]} &
        \frac{\partial \ddot{\mathbf s} (\Psi)}{\partial \phi_i[l_p]}
    \end{bmatrix}
$. Since the maximum likelihood estimates satisfy the zero-gradient condition
$ \nabla_{\Psi}\mathcal J (\Psi) = \mathbf 0$, and by noting that $\frac{\partial \ddot{\mathbf s} (\Psi)}{\partial f_{D,i}}
    =
    -j2\pi(l_p-1)T_s\,\ddot{\mathbf s} (\Psi)$, we obtain
\begin{equation}
    \frac{\partial \mathcal J (\Psi)}{\partial f_{D,i}}
    =
    -2\,\Re\!\left\{
        -j2\pi(l_p-1)T_s\,
        \ddot{\mathbf s} (\Psi)^H(\hat{\mathbf h}[l_\mathrm{p}]-\ddot{\mathbf s} (\Psi))
    \right\}.
\end{equation}

Using the fact that $\Re\{-jz\} = \Im\{z\}$ for a complex-valued $z$, and noting that $\ddot{\mathbf s} (\Psi)^H\ddot{\mathbf s} (\Psi)$ is
real, the zero-gradient condition becomes $ \Im\{\ddot{\mathbf s} (\Psi)^H \hat{\mathbf h}[l_\mathrm{p}]\} = 0$, and thus the Doppler frequency estimate can be found as
\begin{equation}
\hat{f}_{D,i}\!\left[l_\mathrm{p}\right]\! =\!\frac{1}{2\pi (l_\mathrm{p}\!-\!1)T_{s}}\arg\!\! \left( \!
\mathbf{\hat{h}}_{i}^{H}\left[l_\mathrm{p}\right] \mathbf{a}\!\left(\!
\theta _{i}\left[l_\mathrm{p}\right] ,\phi _{i}\left[l_\mathrm{p}\right]
\right) + k\pi \!\right)\!,\!\label{e-fd-est}
\end{equation}
where $\arg(\cdot)$ denotes the argument or the angle of a complex number, $l_\mathrm{p}\neq1$, and $k\in\mathbb{Z}$. It should be mentioned that the Doppler information cannot be extracted from the first arriving pilot signal since the Doppler shift is relative and the first pilot signal in the data sequence is considered to be the reference. 
Similarly, by evaluating the partial derivative with respect to $d_{i}$ and
setting it to $0$, i.e., $\frac{\partial \mathcal J (\Psi)}{\partial d_i}$, the estimated distance can be obtained in a closed form as a
function of $f_{D,i}$, $\theta _{i}\left[l_\mathrm{p}\right] $ and $\phi _{i}\left[l_\mathrm{p}\right] $. The final solution can be expressed as
\begin{equation}
\hat{d}_{i}\!\left[l_\mathrm{p}\right] \!=\!\left\{\!\! \frac{G_{%
\mathrm{T}}G_{\mathrm{R}}\left( \frac{c}{4\pi f_{c}}\right)^2 {d_{0}^{\kappa }\mathcal{N}^{2}}}{{{\Re%
}^{2}\!\left( \!\mathbf{\hat{h}}_{i}^{H}\left[l_\mathrm{p}\right] \mathbf{a}%
\left( \theta _{i}\left[l_\mathrm{p}\right] ,\phi _{i}\left[l_\mathrm{p}\right] \right)\mathrm{e}^{-j2\pi f_{D,i}(l_\mathrm{p}-1)T_{s}}\!
\right) }}\!\!\right\} ^{\!\frac{1}{\kappa }}  \label{e-dis-est}
\end{equation}
By evaluating the partial derivatives with respect to $\theta_i[l_\mathrm{p}]$ and $\phi_i[l_\mathrm{p}]$, the following zero-gradient conditions are obtained.
\begin{equation}
\label{eq:angle_theta}
    \Re\!\left\{
        \left(\frac{\partial \mathbf a(\theta_i[l_\mathrm{p}],\phi_i[l_\mathrm{p}]}{\partial \theta_i[l_\mathrm{p}]}\right)^H
        (\hat{\mathbf h}[l_\mathrm{p}]-\ddot{\mathbf s} (\Psi))
    \right\}
    = 0
\end{equation}
\begin{equation}
\label{eq:angle_phi}
    \Re\!\left\{
        \left(\frac{\partial \mathbf a(\theta_i[l_\mathrm{p}],\phi_i[l_\mathrm{p}]}{\partial \phi_i[l_\mathrm{p}]}\right)^H
        (\hat{\mathbf h}[l_\mathrm{p}]-\ddot{\mathbf s} (\Psi))
    \right\}
    = 0
\end{equation}
These equations are nonlinear and coupled in
$(\theta_i[l_\mathrm{p}],\phi_i[l_\mathrm{p}])$ due to the exponential array manifold structure.
Consequently, the estimates of $\theta _{i}\left[l_\mathrm{p}\right] $ and $\phi _{i}%
\left[l_\mathrm{p}\right] $ cannot be found in closed form. Thus, $\hat{d}_{i}\left[
m,t=1\right] $ and $\hat{f}_{D,i}\left[l_\mathrm{p}\right] $ are substituted in (%
\ref{e-mleee}) and then $\theta _{i}\left[l_\mathrm{p}\right] $ and $\phi _{i}\left[
l_\mathrm{p}\right] $ can be found using optimization methods such as particle swarm
optimization. It is worth mentioning that (\ref{e-mleee}), (\ref{e-fd-est}) and (\ref{e-dis-est}) should
be solved iteratively and alternatively to obtain the final
estimates. Algorithm \ref{alg:MLE} summarizes the entire process, referred to as Alternating Optimization with MLE for $\theta_i$ and $\phi_i$ and closed-form solutions for $d_i$ and $f_{D,i}$ (AO-MLE-CF).\vspace{-0.1in}


\begin{algorithm}
\caption{AO-MLE-CF based estimation of mobility and location parameters}\label{alg:MLE}
\KwData{Received pilot vector $\mathbf{y}_{i,\mathrm{p}}[m,1]$, $\sigma_h^2$, $\sigma_n^2$,$c$,$f_c$}
\KwResult{$\hat{d}_{i}[l_\mathrm{p}],\hat{f}_{D,i}[l_\mathrm{p}],\hat{\theta}_{i}[l_\mathrm{p}],\hat{\phi}_{i}[l_\mathrm{p}]$}

Estimate channel vector using:
\[
\mathbf{\hat{h}}_{i}\left[ m,1\right] =\frac{1}{\sqrt{p_{i,\mathrm{p}}}}%
\mathbf{y}_{i,\mathrm{p}}\left[ m,1\right]
\]

Initialize $\hat{d}_{i}[l_\mathrm{p}],\hat{f}_{D,i}[l_\mathrm{p}],\hat{\theta}_{i}[l_\mathrm{p}],\hat{\phi}_{i}[l_\mathrm{p}]$\;

\For{$k=1$ \KwTo $i_\mathrm{max}$}{
    
    Update $\hat{d}_{i}[l_\mathrm{p}]$ using the closed-form expression given in \eqref{e-dis-est}

    \If{$l_\mathrm{p}\neq 1$}{
      Update $\hat{f}_{D,i}[l_\mathrm{p}]$ using the closed-form in \eqref{e-fd-est}
      }

    Compute the MLE cost function in \eqref{e-mleee}

    Update $\hat{\theta}_{i}[l_\mathrm{p}]$ and $\hat{\phi}_{i}[l_\mathrm{p}]$ by minimizing the argument in \eqref{e-mleee} using MATLAB \texttt{fmincon} with SQP and a tolerance of $10^{-6}$\;
}
\end{algorithm}

{Interestingly, as can be observed from (\ref{e-h-hat}), the channel estimates have
a structure similar to that of the received signals in \gls{mimo} radar. Thus, by relying on (%
\ref{e-h-hat}), other localization algorithms such as \gls{music} can be applied to estimate $\theta _{i}\left[l_\mathrm{p}\right] $ and $\phi
_{i}\left[l_\mathrm{p}\right] $ effectively with less complexity than \gls{mle}, which may require higher-complexity optimization tools. In this work, we apply MUSIC to provide a benchmark for the performance of the MLE-based algorithm. For the sake of brevity, we will not include the implementation of \gls{music} since this can be found in the literature \cite{shuai1,10266765}. Since \gls{music} is able to estimate the azimuth and elevation angles only, the closed form expressions in \eqref{e-fd-est} and \eqref{e-dis-est} need to be applied for estimating the Doppler frequency and distance, respectively. The full estimation process is required to be performed in an alternating manner in a way similar to Algorithm \ref{alg:MLE}, and thus the \gls{music}-based algorithm can be referred to AO-MUSIC-CF.} 
\subsection{Cramér--Rao lower bound for pilot-based parameters estimation}
With the aid of the Slepian--Bangs formula for the \gls{fim} of a complex Gaussian model with a deterministic mean and known covariance, the \gls{fim} for our model in \eqref{e-h-hat} can be formulated as
\begin{equation}
\big[\mathbf{F}(\Psi[l_\mathrm{p}])\big]_{k,q} =
2\,\Re \Big\{ 
\frac{\partial \mu^H[l_\mathrm{p}]}{\partial \Psi^k[l_\mathrm{p}]} \,
\Sigma_i^{-1} \,
\frac{\partial \mu[l_\mathrm{p}]}{\partial \Psi^q[l_\mathrm{p}]}
\Big\}\label{e-fim-1}
\end{equation}
where $k,q \!\in\! \{\!1,2,3,4\!\}$, $\mu[l_\mathrm{p}]\!=\! \sqrt{\!B_{i}\left(
d_{i},f_{c}\right) }\mathbf{a}\!\left( \theta _{i}\left[  l_\mathrm{p}\right]
,\phi _{i}\left[ l_\mathrm{p}\right] \right) \\
 \times\mathrm{e}^{-j2\pi f_{D,i}(l_\mathrm{p}-1)T_{s}}$, and $\Psi^k[l_\mathrm{p}]$ represents the $k$th element of the vector of the CV's parameters $\Psi[l_\mathrm{p}]$. Thus, the \gls{fim} associated with the CV is $\mathbf{F}(\Psi[l_\mathrm{p}]) \in \mathbb{C}^{4\times4}$. By noting that the effective \gls{awgn} vector in \eqref{e-h-hat} is given by $\mathbf{w}=\mathbf{h}_{i,\mathrm{nLoS}}\left[  m,t=1\right] +\frac{\mathbf{n}_i}{\sqrt{p_{i,\mathrm{p}}}}$, whose elements are \gls{iid} with $0$ mean and a variance of $\sigma_\mathrm{eff}^2=\sigma_h^2 + \sigma_n^2/p_{i,\mathrm{p}}$, the \gls{fim} in \eqref{e-fim-1} can be simplified to
\begin{equation}
\big[\mathbf{F}(\Psi[l_\mathrm{p}])\big]_{k,q} =
\frac{2}{\sigma_\mathrm{err}^2}\,\Re \Big\{ 
\frac{\partial \mu^H[l_\mathrm{p}]}{\partial \Psi^k[l_\mathrm{p}]} \,
\frac{\partial \mu[l_\mathrm{p}]}{\partial \Psi^q[l_\mathrm{p}]}
\Big\}\label{e-fim-2}
\end{equation}
where the individual partial derivatives required to construct the \gls{fim} can be derived as
\begin{align}
\frac{\partial  \mu[l_\mathrm{p}]}{\partial \theta_i[l_\mathrm{p}]} \!&\!=\! \sqrt{\!B_i(d_i,f_c)}  \mathrm{e}^{-j2\pi f_{D,i}(l_\mathrm{p}-1)T_s} \frac{\partial}{\partial \theta_i}  \mathbf{a}\!\left( \theta _{i}\left[l_\mathrm{p}%
\right] ,\phi _{i}\left[l_\mathrm{p}\right] \right)\\
\frac{\partial  \mu[l_\mathrm{p}]}{\partial \phi_i[l_\mathrm{p}]} \!&\!= \!\sqrt{\!B_i(d_i,f_c)} \mathrm{e}^{-j2\pi f_{D,i}(l_\mathrm{p}-1)T_s} \frac{\partial}{\partial \phi_i} \mathbf{a}\!\left( \theta _{i}\left[l_\mathrm{p}%
\right] ,\phi _{i}\left[l_\mathrm{p}\right] \right)   \\
\frac{\partial  \mu[l_\mathrm{p}]}{\partial f_{D,i}[l_\mathrm{p}]} \!&\!= \! -j 2\pi (l_\mathrm{p}-1)T_s \sqrt{\!B_i(d_i,f_c)} \, \mathbf{a}\!\left( \theta _{i}\left[l_\mathrm{p}%
\right] ,\phi _{i}\left[l_\mathrm{p}\right] \right)  \nonumber\\
&\, \times \mathrm{e}^{-j2\pi f_{D,i}(l_\mathrm{p}-1)T_s}\\
\frac{\partial  \mu[l_\mathrm{p}]}{\partial d_i[l_\mathrm{p}]} \!&\!=\! {-\frac{\kappa}{2}\frac{\sqrt{\!B_i(d_i,f_c)}}{d_i}\mathbf{a}\!\left( \theta _{i}\left[l_\mathrm{p}%
\right] ,\phi _{i}\left[l_\mathrm{p}\right] \right) }\mathrm{e}^{-j2\pi f_{D,i}(l_\mathrm{p}-1)T_s}
\end{align}
By using \eqref{e-array-resp-1} and \eqref{e-array-resp-2}, the partial derivatives $\frac{\partial}{\partial \theta_i}  \mathbf{a}\!\left( \theta _{i}\left[l_\mathrm{p}%
\right] ,\phi _{i}\left[l_\mathrm{p}\right] \right)$ and $\frac{\partial}{\partial \phi_i}  \mathbf{a}\!\left( \theta _{i}\left[l_\mathrm{p}%
\right] ,\phi _{i}\left[l_\mathrm{p}\right] \right)$ can be found as

\begin{equation}
    \frac{\partial a_{m,n}}{\partial \theta}
\!=\! -\!j\!\frac{2\pi \! \Delta}{\lambda}\cos\!\theta
\big[(m-1)\cos\!\phi \!+\! (n-1)\sin\phi\big]\,
a_{m,n}(\phi,\theta)
\end{equation}
\begin{equation}
{
\frac{\partial a_{m,n}}{\partial \phi}
\!=\! -\!j\!\frac{2\pi \Delta}{\lambda}\sin\!\theta
\big[\!-\!(m\!-\!1)\sin\!\phi \!+\! (n\!-\!1)\!\cos\!\phi\big]\,
\!a_{m,n}\!(\phi,\theta)
}
\end{equation}
where the subscript $i$ and index $[l_\mathrm{p}]$ have been dropped to simplify the notation.
Thereafter, the \gls{fim} $\mathbf{F}(\Psi_i[l_\mathrm{p}])$ can be constructed, and the Cramér–Rao lower bound for each parameter is obtained from the inverse FIM as
\begin{align}
\mathrm{Var}(\hat{\theta}_i[l_\mathrm{p}]) &\ge [\mathbf{F}^{-1}(\Psi_i[l_\mathrm{p}])]_{1,1},\\
\mathrm{Var}(\hat{\phi}_i[l_\mathrm{p}]) &\ge [\mathbf{F}^{-1}(\Psi_i[l_\mathrm{p}])]_{2,2},\\
\mathrm{Var}(\hat{f}_{D,i}[l_\mathrm{p}]) &\ge [\mathbf{F}^{-1}(\Psi_i[l_\mathrm{p}])]_{3,3},\\
\mathrm{Var}(\hat{d}_i[l_\mathrm{p}]) &\ge [\mathbf{F}^{-1}(\Psi_i[l_\mathrm{p}])]_{4,4}.
\end{align}
\section{PASCAL Data Communication}
\subsection{Data symbols detection}
In typical communication systems, after estimating $\mathbf{\hat{h}}_{i}%
\left[l_\mathrm{p}\right] $, the data detection is carried out to detect the transmitted
symbols in the $m$th frame assuming that the channel is fixed during the frame. It can be shown that the \gls{mld} reduces to \gls{mrc}-based detection given by

\begin{equation}
\hat{s}_{i,\mathrm{d}}\left[ l\right] =\underset{\mathcal{S}_{i,\mathrm{d}}}{\arg\max}\frac{1}{\sqrt{p_{i,\mathrm{d}}}}\frac{\mathbf{\hat{h}}%
_{i}^{H}\left[l_\mathrm{p}\right]}{\left\Vert 
\mathbf{\hat{h}}_{i}^{H}\left[l_\mathrm{p}\right] \right\Vert ^{2}} \mathbf{y}_{i,\mathrm{d}}\left[ l\right],\forall l \neq l_\mathrm{p} \label{e-mrc}
\end{equation}%
where $\mathcal{S}_{i,\mathrm{d}}$ is the set of all possible symbols in the constellation. It should be noted that in addition to the estimation errors in $\mathbf{%
\hat{h}}_{i}^{H}\left[l_\mathrm{p}\right] $, the quality of the detector is affected
by outdated channel estimates, which introduce additional errors, especially when
CVs are moving at high speeds.

\subsection{Performance analysis of \gls{pascal}-\gls{ser}: \gls{psk} signalling}
By referring to the \gls{mrc}-based detector given in \eqref{e-mrc}, the decision variable can be defined as
 
\begin{equation}
\mathbf{\ddot{y}}_{i,\mathrm{d}}\left[ l\right] =\frac{1}{\sqrt{p_{i,%
\mathrm{d}}}}\frac{1}{\left\Vert \mathbf{\hat{h}}_{i}^{H}\left[ l_\mathrm{p}\right]
\right\Vert ^{2}}\mathbf{\hat{h}}_{i}^{H}\left[ l_\mathrm{p}\right] \mathbf{y}_{i,%
\mathrm{d}}\left[ l\right],
\end{equation}
which, by substituting $\mathbf{{y}}_{i,\mathrm{d}}\left[ l\right] $ from \eqref{E-y-1}, and noting that the term $\frac{1}{\sqrt{p_{i,%
\mathrm{d}}}}\frac{1}{\left\Vert \mathbf{\hat{h}}_{i}^{H}\left[ l_\mathrm{p}\right]
\right\Vert ^{2}}$ does not affect the phase of the \gls{psk} symbols, can be expressed as
\begin{equation}
\mathbf{\ddot{y}}_{i,\mathrm{d}}\left[ l\right] 
= \sqrt{p_{i,\mathrm{d}}}%
\left\vert \mathbf{\hat{h}}_{i}^{H}\left[ l_\mathrm{p}\right] \mathbf{h}_{i}\left[ l%
\right] \right\vert \mathrm{e}^{j\psi }s_{i,\mathrm{d}}\left[ l\right] 
+\mathbf{\hat{h}}_{i}^{H}\left[ l_\mathrm{p}\right] \mathbf{n}_{i}\left[ l\right] \label{e-y-psi}
\end{equation}
where $\psi=\arg\left(\mathbf{\hat{h}}_{i}^{H}\left[ l_\mathrm{p}\right] \mathbf{h}_{i}\left[ l%
\right]\right)$, which is independent of the equivalent channel envelope $\left\vert \mathbf{\hat{h}}_{i}^{H}\left[ l_\mathrm{p}\right] \mathbf{h}_{i}\left[ l%
\right] \right\vert$. It can be observed from \eqref{e-y-psi} that non-ideal channel compensation, which arises due to imperfect channel estimation and channel outdate caused by mobility, introduces unwanted phase error represented by $\psi$ that degrades the quality of the received signal. To study the impact of $\psi $, note that it imposes a rotation on the signal constellation, which may severely impact the
error rate at the receiver. 

For binary PSK (BPSK), since the transmitted symbols are real-valued, the
receiver can extract the real part of $\mathbf{\ddot{y}}_{i,\mathrm{d}}%
\left[ l\right] $ and then complete the detection process by comparing $\Re\left(\mathbf{\ddot{y}}_{i,\mathrm{d}}%
\left[ l\right] \right)$ with a threshold that is typically $0$ for equal-likelihood symbols, i.e., $\hat s_{i,\mathrm{d}}\left[ l\right] =\mathrm{sign} \left(\Re\left(\mathbf{\ddot{y}}_{i,\mathrm{d}}%
\left[ l\right] \right)\right)$ with $\mathrm{sign}\left( \cdot \right)$ representing the signum function. Therefore,
the conditional \gls{ser} in this case, which is equivalent to the bit error rate of a binary modulation,
is given by
\begin{equation}
P_{s,\mathrm{BPSK}|\psi ,\mathbf{\hat{h}}_{i}^{H}\left[ l_\mathrm{p}\right] ,\mathbf{h}%
_{i}\left[ l\right] }=Q\left( \sqrt{\frac{2p_{i,\mathrm{d}}\gamma[l]}{\sigma
_{n}^{2}}}\cos \psi \right)
\end{equation}
where $\gamma[l]\triangleq\frac{\left\vert \mathbf{\hat{h}}_{i}^{H}%
\left[ l_\mathrm{p}\right] \mathbf{h}_{i}\left[ l\right] \right\vert ^{2}}{\left\Vert \mathbf{\hat{h}}_{i}^{H}\left[ l_\mathrm{p}\right] \right\Vert ^{2}}%
$.

For standard $M$--ary PSK (MPSK), adjacent symbols are separated by an angle of \(2\pi/M\), and the decision boundaries are located halfway between adjacent symbols, i.e., $ \pm \frac{\pi}{M}$. By taking this circular phase into account, the angular distance from the symbol to the nearest boundary is reduced to $
\Delta = \frac{\pi}{M} - |\psi|$, provided that $ |\psi| < \frac{\pi}{M}$ which is a valid assumption at significant \gls{snr}, i.e., the performance is dominated by errors and the communication system is severely degraded when $ |\psi| > \frac{\pi}{M}$. Accordingly, the decision boundaries are located halfway between adjacent symbols, and thus the Euclidean distance from the symbol point to the nearest boundary is reduced to $
d_{\text{proj}} =  2\sqrt{p_{i,\mathrm{d}} \gamma[l]} \sin\!\left(\frac{\pi}{M} - |\psi|\right)$. Consequently,
\begin{eqnarray}
P_{s,\mathrm{MPSK}|\psi ,\mathbf{\hat{h}}_{i}^{H}\left[ l_\mathrm{p}\right] ,\mathbf{h}%
_{i}\left[ l\right] } 
&=&
\Pr\big(|n| > d_{\text{proj}}/2\big)\notag\\
&=& Q\!\left(\frac{d_{\text{proj}}/2}{\sqrt{\sigma
_{n}^{2}/2}}\right).
\end{eqnarray}
which, after substituting for $d_\mathrm{proj}$, can be expressed as
\begin{equation}
P_{s,\mathrm{MPSK}|\psi ,\mathbf{\hat{h}}_{i}^{H}\left[ l_\mathrm{p}\right] ,\mathbf{h}%
_{i}\left[ l\right] }\!=\!Q\!\left( \!\sqrt{\frac{2p_{i,\mathrm{d}}\gamma[l]}{\sigma
_{n}^{2}} }\sin \!\left(\! \frac{%
\pi }{M}-\left\vert \psi \right\vert \!\right)\!\right)\label{e-ser-mpsk-cond}
\end{equation}
For a significant number of antennas, \gls{clt} can be applied to approximate the distribution of ${ \mathbf{\hat{h}}_{i}^{H}%
\left[ l_\mathrm{p}\right] \mathbf{h}_{i}\left[ l\right] }$. The resulting complex Gaussian approximation exhibits independent but not identically distributed real and imaginary parts. Thus, the distribution of $\psi$ can be approximated as von Mises, also known as circular Gaussian and Tikhonov distribution. However, the \gls{pdf} of $\gamma[l]$ is a ratio of dependent non-central Chi-squared random variables which does not have a simple and known closed-form solution. It is worth noting that although the derived solution in \cite{BRODA20091264} considers the ratio of correlated non-central chi-squared random variables, the solution is not traceable and applies to quadratic-form ratios, which is not the case in our work.  Therefore, we provide an approximation to the \gls{ser} by utilizing the autoregressive model introduced in Sec. \ref{sec-autoreg}. Accordingly,  by substituting \eqref{E-ARM-with-Los} in $\mathbf{\hat{h}}_{i}^{H}%
\left[ l_\mathrm{p}\right] \mathbf{h}_{i}\left[ l\right] $, we obtain
\begin{multline}
\mathbf{\hat{h}}_{i}^{H}%
\left[ l_\mathrm{p}\right] \mathbf{h}_{i}\left[ l\right] =\mathbf{\tilde h}_{i,\mathrm{LoS}}^{H}[l_\mathrm{p}]\mathbf{\tilde h}_{i,\mathrm{LoS}}%
\left[ l\right] +\mathbf{h}_{i,\mathrm{nLoS}}^{H}[l_\mathrm{p}]\mathbf{\tilde h}%
_{i,\mathrm{LoS}}\left[ l\right]  \\
+\mathbf{h}_{i,\mathrm{err}}^{H}[l_\mathrm{p}] \mathbf{\tilde h}_{i,\mathrm{LoS}}\left[ l\right]+\rho \left( l\right) \left( \mathbf{\tilde h}_{i,\mathrm{LoS}}^{H}[l_\mathrm{p}]
\mathbf{h}_{i,\mathrm{nLoS}}[l_\mathrm{p}]\right.\\
\left.+\mathbf{h}_{i,\mathrm{nLoS}}^{H}%
[l_\mathrm{p}]\mathbf{h}_{i,\mathrm{nLoS}}[l_\mathrm{p}]+\mathbf{h}_{i,%
\mathrm{err}}^{H}[l_\mathrm{p}]\mathbf{h}_{i,\mathrm{nLoS}}[l_\mathrm{p}]
\right) \\
+\sqrt{1-\rho ^{2}\left( l\right) } \left(  \mathbf{\tilde h}_{i,\mathrm{LoS}}^{H}%
[l_\mathrm{p}]\mathbf{\ddot{n}}_{i,\mathrm{nLoS}}\left[ l\right] +\mathbf{h%
}_{i,\mathrm{nLoS}}^{H}[l_\mathrm{p}]\mathbf{\ddot{n}}_{i,\mathrm{nLoS}}%
\left[ l\right] \right.\\
\left. +\mathbf{h}_{i,\mathrm{err}}^{H}[l_\mathrm{p}]\mathbf{%
\ddot{n}}_{i,\mathrm{nLoS}}\left[ l\right] \right)
\end{multline}
Since $\mathbf{h}_{i,\mathrm{nLoS}}[l]$, $\mathbf{h}_{i,\mathrm{err}}[l]$ and $\mathbf{\ddot{n}}_{i,\mathrm{nLoS}}[l]$ are $0$ mean $\forall l$, by applying the law of large numbers with a significant number of receiving antennas at the \gls{bs}, $\mathbf{\hat{h}}_{i}^{H}%
\left[ l_\mathrm{p}\right] \mathbf{h}_{i}\left[ l\right]$ can be approximated by the following dominant terms
\begin{equation}
 \mathbf{\hat{h}}_{i}^{H}%
\left[ l_\mathrm{p}\right] \mathbf{h}_{i}\left[ l\right]  \approx \mathbf{\tilde h}_{i,\mathrm{LoS}}^{H}[l_\mathrm{p}]\mathbf{\tilde h}_{i,%
\mathrm{LoS}}\left[ l\right] 
+\rho \left( l\right) \left\Vert \mathbf{h}_{i,%
\mathrm{nLoS}}[l_\mathrm{p}]\right\Vert ^{2} \label{e-numer}
\end{equation}
Thus, the numerator of $\gamma[l]$ can be approximated as
\begin{multline}
\!\!\vert \mathbf{\hat{h}}_{i}^{H}%
\!\left[ l_\mathrm{p}\right] \mathbf{h}_{i}\!\left[ l\right]  \vert ^2\!\approx \!\left\vert \mathbf{\tilde h}_{i,\mathrm{LoS}}^{H}%
[l_\mathrm{p}]\mathbf{\tilde h}_{i,\mathrm{LoS}}\left[ l\right] \right\vert
^{2}\!+\!\rho ^{2}\!\!\left( l\right) \left\Vert \mathbf{h}_{i,\mathrm{nLoS}}[l_\mathrm{p}] \right\Vert ^{4}\\
+2\rho \left( l\right) \left\Vert \mathbf{h}_{i,%
\mathrm{nLoS}}[l_\mathrm{p}]\right\Vert ^{2}\Re\left( \mathbf{\tilde h}_{i,%
\mathrm{LoS}}^{H}[l_\mathrm{p}]\mathbf{h}_{i,\mathrm{LoS}}\left[ l\right]
\right)
\end{multline}
Similarly, the denominator of $\gamma[l]$ can be approximated as
\begin{eqnarray}
\left\Vert \mathbf{\hat{h}}_{i}^{H}\![l_\mathrm{p}]\right\Vert ^{2}
\!\!\approx\!\! &\left\Vert \mathbf{\tilde h}_{i,\mathrm{LoS}}^{H}\![l_\mathrm{p}]
\right\Vert ^{2}\!+\!\left\Vert \mathbf{h}_{i,\mathrm{nLoS}}^{H}\![l_\mathrm{p}]
\right\Vert ^{2}\!+\!\left\Vert \mathbf{h}_{i,\mathrm{err}}^{H}\![l_\mathrm{p}]
\right\Vert ^{2} \label{e-denu}
\end{eqnarray}
To simplify the notation, let us define $A=\left\vert \mathbf{\tilde h}_{i,\mathrm{LoS}}^{H}%
[l_\mathrm{p}]\mathbf{\tilde h}_{i,\mathrm{LoS}}\left[ l\right] \right\vert
^{2}$, $B=\Re\left( \mathbf{\tilde h}_{i,%
\mathrm{LoS}}^{H}[l_\mathrm{p}]\mathbf{\tilde h}_{i,\mathrm{LoS}}\left[ l\right]
\right)$, $C=\left\Vert \mathbf{\tilde h}_{i,\mathrm{LoS}}^{H}\![l_\mathrm{p}]
\right\Vert ^{2}$, $\xi=\left\Vert \mathbf{h}_{i,%
\mathrm{nLoS}}[l_\mathrm{p}]\right\Vert ^{2} $ and $\beta=\left\Vert \mathbf{h}_{i,\mathrm{err}}^{H}\![l_\mathrm{p}]
\right\Vert ^{2}$. It should be noted that our objective is to evaluate the \gls{ser} as a function of the CV's parameters, and thus $A$, $B$, $C$ are deterministic values and dependent on $\{\theta _{i}\left[l_\mathrm{p}\right] ,\phi _{i}\left[l_\mathrm{p}\right]
,f_{D,i}\left[l_\mathrm{p}\right] \}$. Therefore, by using these definitions and substituting \eqref{e-numer} and \eqref{e-denu} in \eqref{e-ser-mpsk-cond}, the conditional \gls{ser} for MPSK can be approximated as
\begin{multline}
\!P_{s,\mathrm{MPSK}|\psi ,\mathbf{\hat{h}}_{i}^{H}\left[ l_{\mathrm{p}}%
\right] ,\mathbf{h}_{i}\left[ l\right] }\!\approx\!Q\!\left( \!\sqrt{\frac{2p_{i,%
\mathrm{d}}\left( A+\rho ^{2}(l)\xi ^{2}+2B\rho (l)\xi \right) }{\sigma
_{n}^{2}\left( C+\xi +\beta \right) }}\right.  \\
\left. \times \sin \left( \frac{\pi }{M}-\left\vert \psi \right\vert \right)
\right) 
\end{multline}%
Note that $\mathbf{h}_{i,\mathrm{err}}[l_\mathrm{p}]=\hat {\mathbf{h}}_{i}[l_\mathrm{p}]-{\mathbf{h}}_{i}[l_\mathrm{p}]$, and by using \eqref{E-y-1} and \ref{e-est-ch-1}, we obtain  $\mathbf{h}_{i,\mathrm{err}}[l_\mathrm{p}]=\frac{1}{\sqrt{p_{i,\mathrm{p}}}}{\mathbf{n}}_{i}[l_\mathrm{p}]$ which implies that  $\mathbf{h}_{i,\mathrm{err}}[l_\mathrm{p}]$ is independent of the actual channel ${\mathbf{h}}_{i}[l_\mathrm{p}]$. Therefore, the variables $\xi$ and $\beta$ are independent of each other, and of $\psi$ as shown previously. Consequently, the approximation of the average \gls{ser} can be further simplified to \eqref{e-ser-long-cond-2} on page \pageref{e-ser-long-cond-2}. However, the triple integral does not have a closed-form solution. Nevertheless, as shown in Appendices A and B, by invoking the $Q$-function approximation in \cite{9133171}, and using some numerical integration methods such as Gauss-Chebyshev and Gauss-Laguerre quadrature rules, an accurate approximation for $P_s$ can be found as given in \eqref{e-ser-final2} on page \pageref{e-ser-final2}, where
\begin{itemize}
  \item 
  $\displaystyle 
 \Omega^2(\sigma_\xi^2 u_j,\sigma_\beta^2 t_i)
\!= \!
\frac{2\,p_{i,\mathrm{d}}\!
\left(
\!A + \rho^2(\sigma_\xi^2 u_j)^2 + 2B\rho(\sigma_\xi^2 u_j)
\right)}
{\sigma_n^2 \left( C + \sigma_\xi^2 u_j + \sigma_\beta^2 t_i \right)}.$

  \item $z_k = \cos\!\left(\frac{2k-1}{2N_{\mathrm{GC}}}\pi\right)$, 
  \quad $w_k = \dfrac{\pi}{N_{\mathrm{GC}}}$  (Gauss--Chebyshev nodes and weights).

  \item $M_a$: number of the considered terms in $Q$--function approximation.

  \item $\{t_i,w_i^{\mathrm{GL}}\}$: Gauss--Laguerre nodes and weights for the $\beta$-integral.

  \item $\{u_j,w_j^{\mathrm{GL}}\}$: Gauss--Laguerre nodes and weights for the $\xi$-integral.

  \item $N_{\mathrm{GL1}}, N_{\mathrm{GL2}}, N_{\mathrm{GC}}$: quadrature orders for $\beta$, $\xi$, and $y=|\psi|$, respectively.

  \item $I_0(\kappa)$: modified Bessel function of the first kind (order zero).
  \item $\Gamma(\cdot)$ denotes the Gamma function.
  \item $N_\beta=N_\xi=2\mathcal{N}$ is the degrees of freedom for $\beta$ and $\xi$.
  \item $\sigma_\beta^2=\sigma_n^2/p_{i,\mathrm{p}}$ and $\sigma_\xi^2=\sigma_h^2$ are the variance values of each element in $\mathbf{h}_{i,\mathrm{err}}[l_\mathrm{p}]$ and $\mathbf{h}_{i,\mathrm{nLoS}}[l_\mathrm{p}]$, respectively.
\end{itemize}
\begin{figure*}[tbp]
\begin{equation}
\!P_{s,\mathrm{MPSK}|\psi ,\mathbf{\hat{h}}_{i}^{H}\left[ l_{\mathrm{p}}%
\right] ,\mathbf{h}_{i}\left[ l\right] }\!=\iiint \!Q\!\left( \!\sqrt{\frac{%
2p_{i,\mathrm{d}}\left( A+\rho ^{2}(l)\xi ^{2}+2B\rho (l)\xi \right) }{%
\sigma _{n}^{2}\left( C+\xi +\beta \right) }}\right. 
\left.  \sin \left( \frac{\pi }{M}-\left\vert \psi \right\vert \right)
\right) f_{\left\vert \psi \right\vert }\left( \left\vert \psi \right\vert
\right) f_{\xi }\left( \xi \right) f_{\beta }\left( \beta \right)
d\left\vert \psi \right\vert d\xi d\beta \label{e-ser-long-cond-2}
\end{equation}\vspace{-0.2in}
\end{figure*}

\begin{figure*}[tbp]
\begin{multline}
P_s \!\approx \! \frac{1}{\Gamma(N_\xi/2)\!\Gamma(N_\beta/2)} \,
\!\!\sum_{j=1}^{N_{\mathrm{GL2}}}\! w_j^{\mathrm{GL}}\, u_j^{\,N_\xi/2-1} \;
\!\sum_{m=1}^{M_a}\!\alpha_m
\!\!\sum_{k=1}^{N_{\mathrm{GC}}} \!w_k\,\frac{e^{\kappa z_k}}{\pi I_0(\kappa)}
\!\!\sum_{i=1}^{N_{\mathrm{GL1}}} \!w_i^{\mathrm{GL}}\,t_i^{\,N_\beta/2-1}\\
\times \exp\!\!\Big[-\lambda_m\,\Omega^2(\sigma_\xi^2 u_j,\!\sigma_\beta^2 t_i)\,
\sin^2\!\big(\tfrac{\pi}{M}\!-\!\arccos\! z_k\big)\Big].\label{e-ser-final2}
\end{multline}\vspace{-0.3in}
\end{figure*}

\section{Results}

\begin{figure}[tb]
\centering
\includegraphics[
scale=0.80
]%
{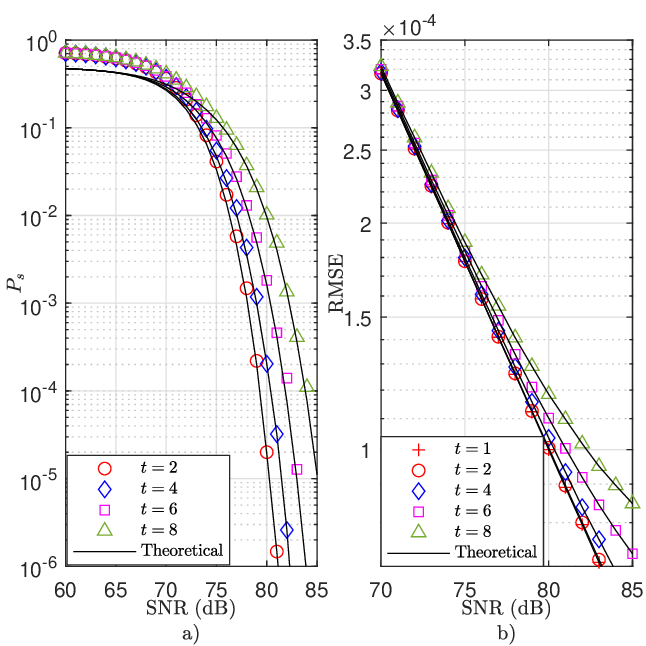}%
\caption{The impact of outdated channel on $P_s$.}
\label{fr1}%
\end{figure}

\begin{figure*}[ht]

\centering
\includegraphics[width=6.8in]{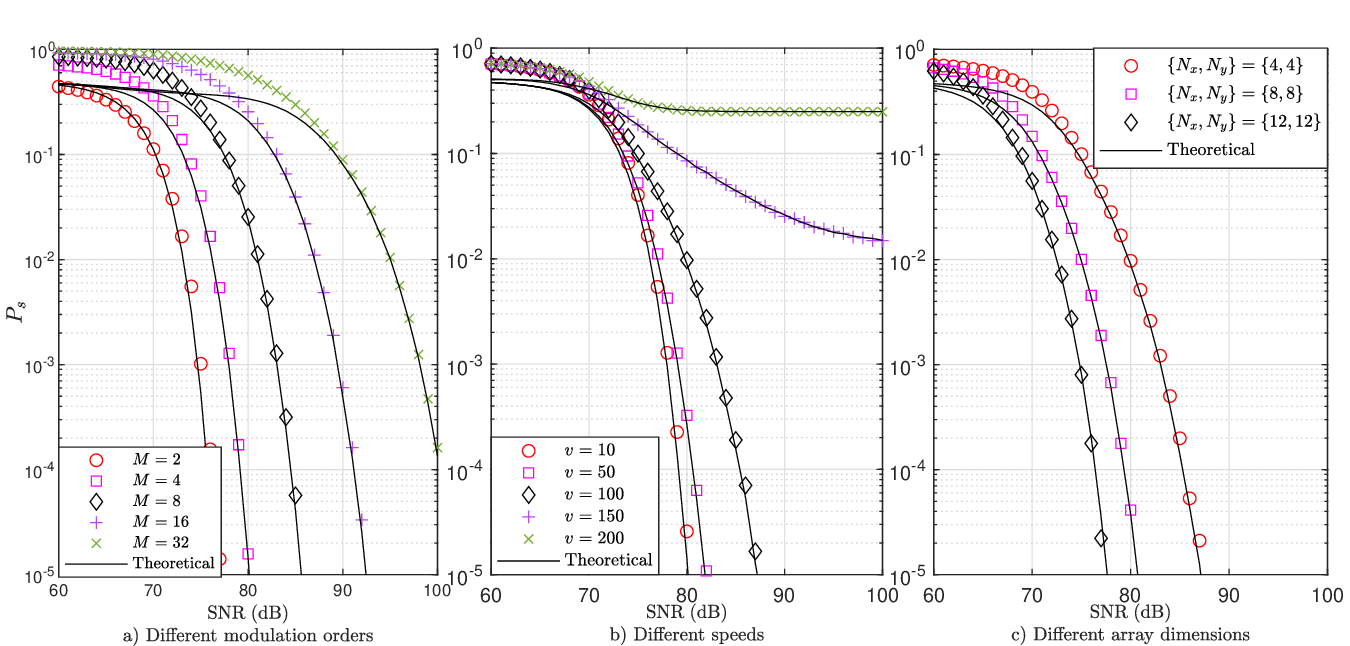}

\caption{SER for various operating condition.}
\label{fr5_6_7}
\vspace{-0.1in}
\end{figure*}

\begin{figure*}[ht]

\centering
\includegraphics[width=6.8in]{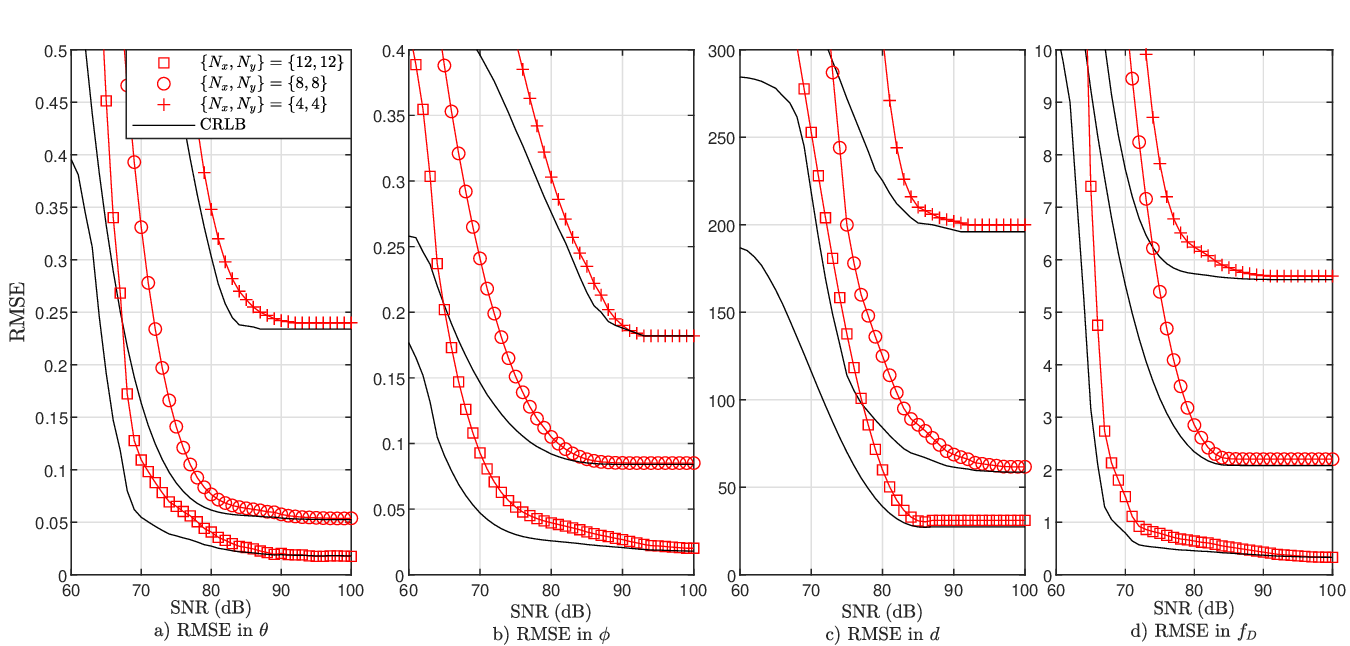}

\caption{The RMSE of the location parameters estimation using AO-MLE-CF algorithm.}
\label{fr2}
\vspace{-0.1in}
\end{figure*}

\begin{figure}[tb]
\centering
\includegraphics[
scale=0.80
]%
{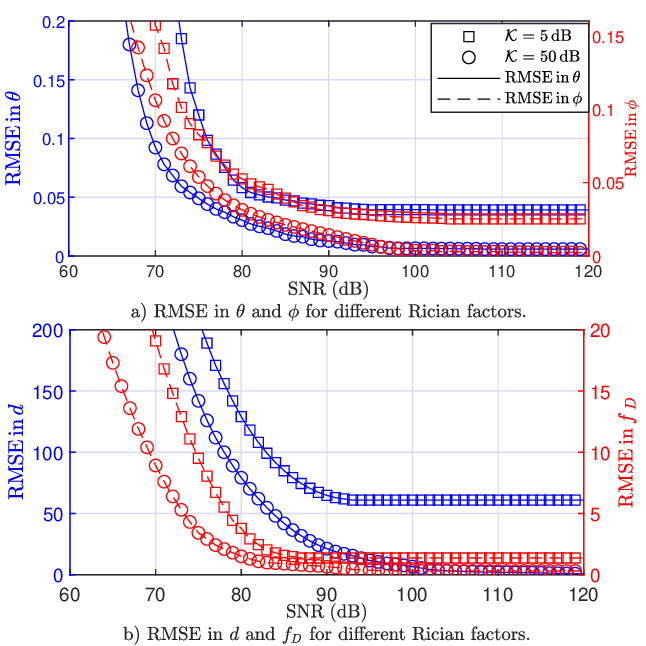}%
\caption{The achievable RMSE using MLE for different Rician factors $\mathcal{K}$.}
\label{fr3}%
\end{figure}

\begin{figure}[tb]
\centering
\includegraphics[
scale=0.80
]%
{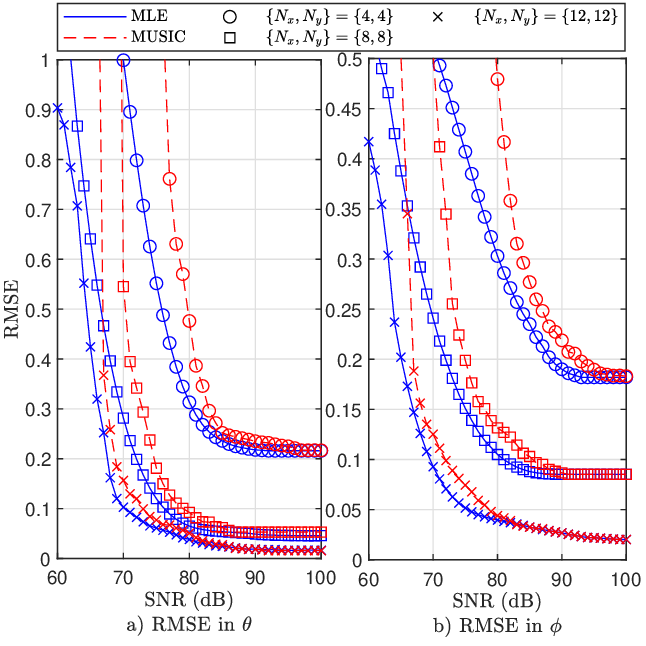}%
\caption{A comparison between MLE and MUSIC.}
\label{fr8}%
\vspace{-.1in}\end{figure}

\begin{figure}[tb]
\centering
\includegraphics[
scale=0.80
]%
{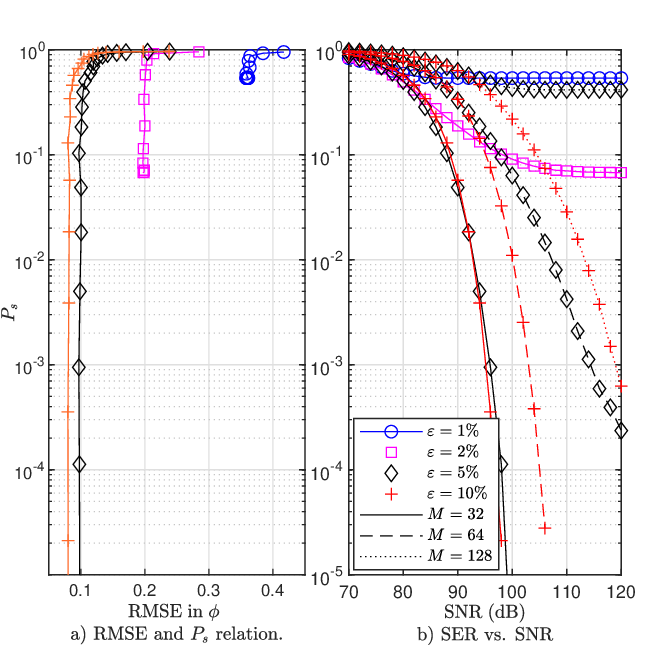}%
\caption{A comparison between MLE and MUSIC.}
\label{fr9}%
\vspace{-.15in}\end{figure}

The simulation results for the \gls{pascal} model introduced in this paper are
presented in this section, where Monte Carlo simulation is considered using $10^6$ Monte Carlo runs. The achievable \gls{ser}, $P_s$, and \gls{rmse} are considered to investigate the system capability. Throughout this  section, the simulation uses a carrier frequency of $f_c = 1\,\text{GHz}$, the path-loss exponent $\kappa$ is set to $2$, and the antenna spacing is set to half the wavelength. Pilots are inserted every $T$ symbols, with uniform power allocation among pilots and symbols. Unless otherwise stated, the Rician factor is $\mathcal{K}=10$ dB, and the variance of the \gls{nlos} random component is computed accordingly, i.e., $\sigma_h^2=\frac{\mathcal{K}}{B_i(d_i,f_c)}$. Furthermore, unless otherwise mentioned, the speed is $100$ km/h, a circular trajectory whose center is above the \gls{bs}, with a height of $h_0=100$ m and a radius of $R=1000\,\text{m}$ is considered, the array size is $4\times4$, and the applied modulation is QPSK. {For Figs.~\ref{fr1}, \ref{fr5_6_7}, \ref{fr2}, \ref{fr3}, and~\ref{fr8}, a symbol rate of $R_s = 10^4$ symbols/s is considered, where such a rate is appropriate for telemetry communication, e.g., control data, sensory readings, etc, while the symbol rate in Fig. \ref{fr9} is $R_s = 5\times10^5$ symbols/s, which is sufficient for narrowband data transmission}. The trajectory obeys the mathematical model given in Table \ref{tab:trajectory}, where $\omega$ is the angular resolution of the \gls{uav}, which can be defined as $\omega=\frac{2\pi}{T_F}$, where $T_F$ is the flight duration, defined as $T_F=N_\mathrm{symb}/R_s$, where $N_\mathrm{symb}$ is the total number of symbols to be transmitted. The average received \gls{snr} is defined as $\mathrm{SNR_r}=\frac{p_{i,\mathrm{d}}E[||\mathbf{h}||^2]  }{\sigma_n^2}$; this is why the range of the transmit SNR in the figures seems high which is basically to compensate for the path loss. AO-MLE-CF algorithm is applied in all figures except Fig. \ref{fr8}, which compares the achievable RMSE performance of MLE and MUSIC.
\begin{table}[ht]
\centering
\caption{Trajectory equations for circular motion around a fixed receiver.}
\begin{tabular}{|c|c|}
\hline
\textbf{Quantity} & \textbf{Equation} \\
\hline
X-position & 
\(
 x(t) = R \cos(\omega t)
\) \\
\hline
Y-position & 
\(
 y(t) = R \sin(\omega t)
\) \\
\hline
Z-position & 
\(
 z(t) = h_0
\) \\
\hline
Distance to receiver & 
\(
d(t) = \sqrt{x(t)^2 + y(t)^2 + z(t)^2}
\) \\
\hline
Elevation angle & 
\(
\theta(t) = \arctan\Big(\frac{h_0}{\sqrt{x(t)^2 + y(t)^2}}\Big)
\) \\
\hline
Azimuth angle & 
\(
\phi(t) = \text{arctan}\big(\frac{y(t)}{ x(t)}\big)
\) \\
\hline
\end{tabular}

\label{tab:trajectory}
\end{table}

Fig. \ref{fr1} illustrates the impact of the outdated channel by plotting the \gls{ser} of various data symbols in the frame. The transmitter moves at $100\,\text{km/h}$, the array size is $4\times4$, $T=10$,
and the modulation scheme is QPSK. As can be observed from Fig. \ref{fr1}(a), the outdated channel has a significant impact on the achievable \gls{ser}. For instance, the \gls{ser} of the first data symbol in the frame, i.e., $t=2$, exhibits the best performance among the others since it has a relatively high correlation with the channel realization at the pilot instance. It can also be noted that as the time separation between the pilot and the detected symbol increases, the performance becomes worse. For example, at $P_s=10^{-5}$, the performance gap between the cases of $t=2$ and $t=4$ is about $1$ dB, whereas it is more than $4.5$ dB when comparing $t=8$ to $t=2$. Fig. \ref{fr1}(b) shows the \gls{rmse} of the channel estimates including the errors induced by outdating. As can be seen from Fig. \ref{fr1}(b), the smaller the time gap between the pilot position, i.e., $t=1$, and the data symbols, the lower the \gls{rmse}, which is reflected in the lower \gls{ser} shown in Fig. \ref{fr1}(a). It can also be observed from Fig. \ref{fr1}(b) that the outdating has a negligible impact on the channel estimates at $t=2$. However, as $t$ increases, the \gls{rmse} increases, causing the last data symbol in the frame to have the worst-quality channel estimate, thus resulting in the worst detection performance. Moreover, the figure confirms the derived formulae for the approximated \gls{ser} and channel estimation \gls{rmse} as the theoretical results match the simulation results. It is noteworthy that the achievable SER and RMSE shown in this figure are subject to all system imperfections including \gls{awgn}, fading, and channel outdating caused by time variations.

Fig. \ref{fr5_6_7} shows the effect of several operating conditions on the achievable SER, i.e., different modulation orders, several speed values, and different array sizes are considered in Figs. \ref{fr5_6_7}(a), \ref{fr5_6_7}(b) and \ref{fr5_6_7}(c), respectively. The frame length is fixed at $T=10$ with one pilot inserted at the beginning of the frame. As can be observed from the figure, the derived approximation for the SER provides an excellent match with the error rates obtained from simulation, except in the very low SNR range, which confirms the accuracy of the derived formula for the typical SNR range. As can be depicted from Fig. \ref{fr5_6_7}(a), as the modulation order increases, the SER increases. For example, a $P_s$ of $10^{-5}$ is achieved by using QPSK at $80$ dB whereas it reaches about $2.5\times 10 ^{-2}$ with 8PSK, and it is more than $10^{-1}$ when $M>8$. It can also be observed from Fig. \ref{fr5_6_7}(b) that the SER becomes worse as the velocity increases. For example, the achievable SER when the speed is as low as $10$ km/h is about $10^{-5}$ at $80$ dB, while it is as high as $0.01$ at the same SNR value when the speed reaches $100$ km/h. As can also be observed, by increasing the speed to $150$ km/h and beyond, the SER suffers from an error floor, whose value increases with the vehicle speed. It is worth noting that the higher the speed, the lower the correlation over time. This leads to significant errors in channel equalization due to considerable outdated channel measurements. Finally, Fig. \ref{fr5_6_7}(c) shows the impact of the array size on the achievable SER, where the larger the array, the better the SER, which may partially compensate for channel errors by providing spatial diversity.

In Fig. \ref{fr2}, we demonstrate the capability of the AO-MLE-CF algorithm, where MLE is applied to estimate $\theta$ and $\phi$ and the closed-form estimators in \eqref{e-dis-est} and \eqref{e-fd-est} are employed for $d$ and $f_D$, respectively, in an iterative manner. The frame length is $T=10$ with one pilot inserted at the beginning of every frame. The results in this figure are obtained for several array dimensions, and the \gls{crlb} is plotted. The number of iterations for AO is fixed at 20, and the Matlab built-in function $\mathrm{fmincon}$ is applied to optimize the maximum likelihood function using the Sequential Quadratic Programming (SQP) algorithm with a maximum of 30 iterations and a convergence tolerance of $10^{-8}$. As can be observed from all subplots, increasing the array size manages to enhance the RMSE; however, under a certain array size, the improvement rate is low at high SNRs, and the RMSE tends to converge to a certain value that is comparable with the \gls{crlb} limit. For instance, Fig. \ref{fr2}(a) shows that for the $12\times  12$ array size scenario, increasing the SNR from $70$ dB to $80$ dB manages to reduce the RMSE in $\theta$ from $0.11$ rad to $0.04$ rad, while increasing it from $80$ dB to $90$ dB reduces the RMSE from $0.02$ rad to $0.018$ rad. As can also be seen, at an SNR of $90$ dB, Figs. \ref{fr2}(a), (b), (c) and (d) show that increasing the array size from $4\times 4$ to $12\times 12$ leads to a reduction in the RMSE of $\theta$, $\phi$, $d$ and $f_D$, respectively, from $0.24$ rad to $0.018$ rad, from $0.19$ rad to $0.027$ rad, from $202$ m to $30$ m, and from $5.7$ Hz to $0.42$ Hz. It is worth noting that the distance estimation error in Fig. \ref{fr2}(d) is significant even at high SNR, which implies that for accurate distance estimation, much larger antenna arrays are required, or better Rician channel conditions are needed, as will be shown in Fig. \ref{fr3}. Furthermore, the figure shows that RMSEs converge with the derived \gls{crlb} at high SNRs. 

Fig. \ref{fr3} shows the impact of the strength of the LoS channel component on the achievable estimation error by plotting the RMSE with different Rician factor values. The operating conditions are similar to the previous figure, except that the array size is fixed at $12\times 12$, the frame length is $T=10$, and the Rician factor is varied. As can be depicted from the figure, the Rician factor has a significant impact on the achievable RMSE, where the higher $\mathcal{K}$ the lower the RMSE in all parameters. This is due to the fact that the \gls{nlos} component acts like a random noise. Thus, by strengthening the LoS, in which the information about the parameters is embedded, more information about the estimated parameters can be extracted. For example, it can be seen from Fig. \ref{fr3}(a) that at an SNR of $100$ dB, the RMSE in $\theta$ decreases from $0.04$ rad to about $0.01$ rad by increasing the Rician factor from $5$ dB to $10$ dB. Moreover, for the same SNR, the RMSE in $d$ is reduced from $60$ m to $5$ m for the same Rician factors. Therefore, to enable accurate distance estimation, a large antenna array and a significant Rician factor are required.

Fig. \ref{fr8} compares the estimation capability of the MLE and MUSIC algorithms under different array dimensions by considering the RMSE in estimating $\theta$ and $\phi$ for given $f_D$ and range $d$ which are assumed to be known in this figure. In this figure, the modulation order is fixed at $M=4$, the frame length is $T=10$, the speed is $100\,\text{km/h}$, and the Rician factor is $10$ dB. The figure shows that MLE provides better performance than MUSIC at low and mid ranges of SNR regardless of the array size. For example, as can be seen from Fig. \ref{fr8}(b), at an SNR of $80$ dB and when $\{N_x,N_y\}=\{4,4\}$, the RMSE in $\phi$ using MLE is about $0.3$ rad, whereas it is about $0.45$ rad using MUSIC. However, the gap between both estimators reduces as the SNR and the array size increase, which leads to converging RMSEs at high SNRs.\vspace{-.01in}

Fig. \ref{fr9} provides a comparison of the achievable PASCAL performance under several values of the pilot-to-frame ratio, where this ratio is defined as $\varepsilon\triangleq \frac{N_\mathrm{p}}{T}\times 100\%$ where $N_p$ is the number of pilots inserted in a frame of a total number of $T$ symbols, with $T=100$ being considered in this figure. Compared to the previous figure, the symbol rate in this figure is $R_s=5\times 10^{5}$ symbols per second, which within the typical range of narrowband data communication. Moreover, the trajectory here is modified to create a tilted, i.e., $y(t)=R\sin(\omega t)\cos(\beta)$, with $\beta=30^o$ being the inclination angle,   circular motion with a periodically oscillating altitude, i.e., $z(t)=h_0+R\sin(\omega t)\sin(\beta)$, to provide more dynamic UAV mobility conditions. The velocity and array size are fixed at $200$ km/h and $\{4,4\}$, respectively, while all other parameters are set similarly to the previous figures. Fig. \ref{fr9}(a) shows the RMSE in $\phi$ vs. the SER, where multiple values of $\varepsilon$ are used. For each curve in this subplot, the SNR is varied and a pair of the achievable $\{\mathrm{RMSE}\;\mathrm{in}\;\phi,P_s\}$ at each SNR value is considered for the plot, where AO-MLE-CF is applied for localization. We plot the RMSE in $\phi$ only because of the limited space, however, similar conclusions are obtained from other estimation parameters.  As can be observed from this subfigure, the pilot percentage has a direct impact on the achievable RMSE and $P_s$, and the RMSE-$P_s$ relation is a win-win, where the better the RMSE, the lower $P_s$. It can be seen from the figure that when $\varepsilon=1\%$ both the RMSE and $P_s$ have minimum values (e.g. error floors) of about $0.36$ rad and $0.5$, respectively, where their values cannot be decreased even with higher SNR; this is why the curve in this case, as well as when $\varepsilon=2\%$ seems to be truncated and does not extend to cover wide ranges of the RMSE and $P_s$. By increasing $\varepsilon$ to $2\%$ the minimum values can be decreased to about $0.2$ and $4.5\times10^{-5}$.  To overcome these error values, $\varepsilon$ needs to be increased. For example, as can be depicted from the figure, the error floor disappears when $\varepsilon=\{5\%,10\%\}$. This enhancement in $P_s$ and RMSE is achieved at the expense of a lower transmission rate due to increasing the number of pilots in a frame, but it is necessary for reliable symbol detection.   As shown in Fig. \ref{fr9}
(b), the transmission rate can be enhanced by increasing the modulation order $M$. For example, the figure shows that the error floor persists when $\varepsilon$ increases from $1\%$ to $2\%$, and stays at high levels even when the modulation order is $M=32$. However, by increasing $\varepsilon$ to $10\%$ the error floor is completely removed even with higher modulation orders such as $64$ and $128$, which implies that even though the number of data symbols per a frame is reduced to $90\%$, the bit rate is boosted while achieving much lower SER compared to the cases of $\varepsilon=\{1\%,2\%\}$. The observations in this figure imply that the pilot percentage and the modulation order are design parameters that should be carefully selected to guarantee a balance between data rate and SER, as well as a desirable RMSE.
\color{black}
\section{Conclusion}
In this paper, we investigated a novel system model, named PASCAL, to enable efficient simultaneous localization and communication by exploiting embedded pilot symbols within the communication signal. We explored the channel estimation, data detection, and location parameter estimation under a variety of operating conditions, presenting and analyzing several algorithms. The channel was modeled as Rician, capturing the effects of random \gls{nlos} components on the direct path, subject to Jakes’ model, which captures temporal channel correlations induced by mobility. Additionally, a first-order autoregressive model was employed to approximate errors arising from outdated channel estimates.

Our findings revealed that, with suitable SNR and antenna array configurations, pilot signals can effectively support localization in addition to conventional channel estimation. Notably, unlike traditional \gls{isac} approaches, PASCAL achieves this without requiring any additional hardware or signaling design, highlighting its potential as a practical and efficient framework for integrated localization and communication systems. In addition, it was shown that the proposed  MLE outperforms  MUSIC for a wide range of operating SNR, especially at low- and mid-ranges. For high speed scenarios, the system suffered from significantly high SER, which converged to an error floor due to outdated channel conditions, which can be solved by increasing the frequency of the pilot signals. Moreover, our results demonstrated an excellent agreement between the analytical SER and simulation outcomes, with the RMSE converging to the derived \gls{crlb} at high SNRs. \vspace{-.05in}

\section{Appendix A: Averaging over $|\psi|$\vspace{-.05in}}
Let \(\psi \sim \text{von Mises}(0,\kappa) \) with the \gls{pdf}
\begin{equation}
    f_\Psi(\psi) = \frac{e^{\kappa \cos \psi}}{2\pi I_0(\kappa)}, \quad x \in [-\pi,\pi),
\end{equation}
where \( I_0(\kappa) \) is the modified Bessel function of the first kind. Since the \gls{pdf} \( f_Y(y) \) is an even function, the \gls{pdf} of \( y \triangleq|\psi| \) is
\begin{equation}
    f_Y(y) = f_\psi(y) + f_\psi(-y) = \frac{e^{\kappa \cos y}}{\pi I_0(\kappa)}, \quad 0 \le y < \pi
\end{equation}
Therefore, by considering the integral in \eqref{e-ser-long-cond-2} with respect to $y=|\psi|$, the \gls{ser} equation reduces to
\begin{equation}
I_y=\int_ {0}^{\infty} \!Q\!\left(\Gamma(\xi,\beta)
  \sin \left( \frac{\pi }{M}-y \right)
\right) \frac{e^{\kappa \cos y}}{\pi I_0(\kappa)} 
dy \label{e-ser-long-cond-Iy}
\end{equation}
where 
\begin{equation}
\Omega(\xi,\beta)=\sqrt{\frac{2p_{i,%
\mathrm{d}}\left( A+\rho ^{2}(l)\xi ^{2}+2B\rho (l)\xi \right) }{\sigma
_{n}^{2}\left( C+\xi +\beta \right) }}.
\end{equation}
Thereafter, the $Q$-function approximation with a finite sum is invoked to simplify the integration \cite{9133171}, which yields
\begin{equation}
I_y \!\approx \!
\sum_{m=1}^{M_a} \alpha_m \!\int_0^{\pi} \!
\!\exp\!\Big[\!-\lambda_m \, \Omega^2(\xi,\beta) \, \sin^2\!\Big(\frac{\pi}{M} - y\Big)\!\Big] 
\frac{e^{\kappa \cos y}}{\pi I_0(\kappa)} \, dy
\label{eq:Iy_tanash}
\end{equation}
Using integration by substitution with $z = \cos y$, we obtain
\begin{equation}
I_y \!\approx \!
\sum_{m=1}^{M_a}\!\! \alpha_m\! \int_{-1}^{1} \!\!
\frac{\!\exp\!\Big[\!-\lambda_m \!\Omega^2\!(\!\xi,\beta)\! \sin^2\!(\pi/M \!- \!\arccos z)\!\Big] \! e^{\kappa z}}
{\pi I_0(\kappa) \sqrt{1-z^2}}  dz,
\label{eq:Iy_cos_sub}
\end{equation}
which can be efficiently solved using the Gauss–Chebyshev quadrature rule \cite{Abramowitz1964}, as
\begin{equation}
I_y\! \approx \!\sum_{m=1}^{M_a}\!\! \alpha_m \!\!
\sum_{k=1}^{N_\mathrm{GC}} \!\!w_k \, 
\frac{\exp\!\Big[\!-\!\lambda_m \Omega^2\!(\!\xi,\beta) \!\sin^2\!(\frac{\pi}{M}\! -\! \arccos z_k)\Big]\! e^{\kappa z_k}}
{\pi I_0(\kappa)}
\label{eq:Iy_chebyshev_sum}
\end{equation}
where $z_k = \cos\frac{2k-1}{2N} \pi$, $N_\mathrm{GC}$ is the number of Gauss–Chebyshev quadrature nodes, and $w_k = \frac{\pi}{N}$. Consequently, the integral in \eqref{e-ser-long-cond-2} is reduced to 
\begin{equation}
\!P_{s,\mathrm{MPSK}|\psi ,\mathbf{\hat{h}}_{i}^{H}\left[ l_{\mathrm{p}}%
\right] ,\mathbf{h}_{i}\left[ l\right] }\!=\iint I_y f_{\beta }\left( \beta \right)f_{\xi }\left( \xi \right) 
d\beta d\xi  \label{e-ser-long-cond-3}
\end{equation}

\section*{Appendix B: Averaging over $\beta$ and $\xi$}

Next, we evaluate the integral with respect to $\beta$, which can be written as
\begin{equation}
I_\beta(\xi) = \int_0^\infty I_y(\xi,\beta) f_\beta(\beta) \, d\beta,
\end{equation}
where $\beta$ is a scaled Chi-squared random variable for which the \gls{pdf} is given by
\begin{equation}
f_\beta(\beta) = \frac{\beta^{N_\beta/2-1} e^{-\beta/\sigma_\beta^2}}{(\sigma_\beta^2)^{N_\beta/2} \Gamma(N_\beta/2)}, \quad \beta \ge 0,
\end{equation}
where $N_\beta=2\mathcal{N}$ is the degrees of freedom of $\beta$ and $\sigma_\beta^2=\sigma_n^2/p_{i,\mathrm{p}}$ is the variance of each element in $\mathbf{h}_{i,\mathrm{err}}[l_\mathrm{p}]$. By substituting the corresponding expressions of $I_y$ and $\Omega(\xi,\beta)$, we obtain the equation \eqref{e-I-beta1} provided on page \pageref{e-I-beta1}.
\begin{figure*}[tbp]
\begin{multline}
I_\beta(\xi) = \sum_{m=1}^{M_a} \alpha_m \sum_{k=1}^{N_\mathrm{GC}} w_k \frac{e^{\kappa z_k}}{\pi I_0(\kappa) (\sigma_\beta^2)^{N_\beta/2} \Gamma(N_\beta/2)}
\int_0^\infty \beta^{N_\beta/2-1} \\
\times \exp\Bigg[-\lambda_m \frac{2 p_{i,d} (A + \rho^2 \xi^2 + 2 B \rho \xi)}{\sigma_n^2 (C + \xi + \beta)} \sin^2(\pi/M - \arccos z_k) - \frac{\beta}{\sigma_\beta^2} \Bigg] d\beta.
\label{e-I-beta1}
\end{multline}\vspace{-.2in}
\end{figure*}

Let $\beta = \sigma_\beta^2 t$, where $t \sim \text{Gamma}(N_\beta/2,1)$, and by noting that the resulting expression takes the form of a Gauss-Laguerre quadrature integral \cite{Concus1963}, thus $I_\beta(\xi)$ can be found as \eqref{e-I-beta2} on page \pageref{e-I-beta2},
\begin{figure*}[tbp]
\begin{equation}
I_\beta(\xi) \approx \frac{1}{\Gamma(N_\beta/2)} \sum_{m=1}^{M_a} \alpha_m \sum_{k=1}^{N_\mathrm{GC}} w_k \frac{e^{\kappa z_k}}{\pi I_0(\kappa)}
\sum_{i=1}^{N_\mathrm{GL1}} w_i^{GL} t_i^{N_\beta/2-1} 
\exp\Big[-\lambda_m \Omega^2(\xi,\sigma_\beta^2 t_i) \sin^2(\pi/M - \arccos z_k) \Big]
\label{e-I-beta2}
\end{equation}\vspace{-.2in}
\end{figure*}
where $t_i$ and $w_i^\mathrm{GL}$ are the Gauss–Laguerre nodes and weights of order $N_\mathrm{GL1}$, and $\Omega^2(\xi,\sigma_\beta^2 t_i) $ is 
\begin{equation}
\Omega^2(\xi,\sigma_\beta^2 t_i) = 
\frac{2 p_{i,d} (A + \rho^2 \xi^2 + 2 B \rho \xi)}{\sigma_n^2 (C + \xi + \sigma_\beta^2 t_i)}.
\end{equation}

Finally, using a similar approach to solve the third integral with respect to $\xi$, the average \gls{ser} can be approximated as \eqref{e-final-0bv} on page \pageref{e-final-0bv},
\begin{figure*}[tbp]
\begin{multline}
P_s \approx \; \frac{1}{\Gamma(N_\xi/2)\,\Gamma(N_\beta/2)} \,
\sum_{j=1}^{N_{\mathrm{GL2}}} w_j^{\mathrm{GL}}\, u_j^{\,N_\xi/2-1} \;
\sum_{m=1}^{M_a}\alpha_m
\sum_{k=1}^{N_{\mathrm{GC}}} w_k\,\frac{e^{\kappa z_k}}{\pi I_0(\kappa)}\\
\times\sum_{i=1}^{N_{\mathrm{GL1}}} w_i^{\mathrm{GL}}\,t_i^{\,N_\beta/2-1}
\exp\!\Big[-\lambda_m\,\Omega^2(\sigma_\xi^2 u_j,\sigma_\beta^2 t_i)\,
\sin^2\!\big(\tfrac{\pi}{M}-\arccos z_k\big)\Big].\label{e-final-0bv}
\end{multline}\vspace{-0.3in}
\end{figure*}
where $N_\xi=2\mathcal{N}$ is the degrees of freedom of $\xi$ and $\sigma_\xi^2=\sigma_h^2$ which is the variance of element in $\mathbf{h}_{i,\mathrm{nLoS}}[l_\mathrm{p}]$. The variable $\Omega^2(\sigma_\xi^2 u_j,\sigma_\beta^2 t_i)$ is given by
\begin{equation}
    \Omega^2(\sigma_\xi^2 u_j,\sigma_\beta^2 t_i)
\!= \!
\frac{2\,p_{i,\mathrm{d}}\!
\left(\!
A \!+\! \rho^2(\sigma_\xi^2 u_j)^2 \!+\! 2B\rho(\sigma_\xi^2 u_j)\!
\right)}
{\sigma_n^2 \left( C + \sigma_\xi^2 u_j + \sigma_\beta^2 t_i \right)}.
\end{equation}

\bibliographystyle{IEEEtran}
\bibliography{Ref.bib}
\end{document}